\documentclass[fleqn,usenatbib]{mnras}
\pdfoutput=1

\usepackage{newtxtext,newtxmath}
\usepackage{bm}

\usepackage[T1]{fontenc}
\usepackage{ae,aecompl}

\usepackage{graphicx}	
\usepackage{amsmath}	
\usepackage{amssymb}	

\newcommand{\bmk}{\bm{k}}
\newcommand{\bmv}{\bm{v}}
\newcommand{\bmx}{\bm{x}}
\newcommand{\bmr}{\bm{r}}
\newcommand{\bmd}{\bm{d}}
\newcommand{\bms}{\bm{s}}
\newcommand{\bmR}{\bm{R}}
\newcommand{\bmF}{\bm{F}}
\newcommand{\bmp}{\bm{\psi}}
\newcommand{\bmeps}{\bm{\epsilon}}

\newcommand{\mr}{\mathrm}

\title[Reconstruction with velocities]{Reconstruction with velocities}

\author[H.-M. Zhu et al.]{
Hong-Ming Zhu,$^{1,2}$\thanks{E-mail: hmzhu@berkeley.edu}
Martin White,$^{1,2,3}$
Simone Ferraro,$^{2,1}$
and Emmanuel Schaan$^{2,1}$
\\
$^{1}$Berkeley Center for Cosmological Physics and Department of Physics, University of California, Berkeley, CA 94720, USA\\
$^{2}$Lawrence Berkeley National Laboratory, 1 Cyclotron Road, Berkeley, CA 94720, USA\\
$^{3}$Department of Astronomy, University of California, Berkeley, CA 94720, USA
}

\date{Accepted XXX. Received YYY; in original form ZZZ}

\pubyear{2020}

\begin{document}
\label{firstpage}
\pagerange{\pageref{firstpage}--\pageref{lastpage}}
\maketitle

\begin{abstract}
Reconstruction is becoming a crucial procedure of galaxy clustering analysis for future spectroscopic redshift surveys to obtain subpercent level measurement of the baryon acoustic oscillation scale.
Most reconstruction algorithms rely on an estimation of the displacement field from the observed galaxy distribution.
However, the displacement reconstruction degrades near the survey boundary due to incomplete data and the boundary effects extend to $\sim100\,\mathrm{Mpc}/h$ within the interior of the survey volume.
We study the possibility of using radial velocities measured from the cosmic microwave background observation through the kinematic Sunyaev-Zel'dovich effect to improve performance near the boundary.
We find that the boundary effect can be reduced to $\sim30-40\,\mathrm{Mpc}/h$ with the velocity information from Simons Observatory.
This is especially helpful for dense low redshift surveys where the volume is relatively small and a large fraction of total volume is affected by the boundary.
\end{abstract}


\section{Introduction}

Precision measurements of the baryon acoustic oscillations (BAO) can constrain dark energy models and modified gravity theories on cosmological scales \citep[e.g.][]{2013PhR...530...87W,2018LRR....21....2A}.
Future stage IV galaxy surveys plan to measure the BAO scale to subpercent precision over a wide range of redshifts \citep[e.g.][]{2016arXiv161100036D,2014PASJ...66R...1T,2011arXiv1110.3193L,2018LRR....21....2A,2009arXiv0912.0201L,2016arXiv160407626D}.
However, nonlinear clustering due to gravitational instabilities smears the linear BAO feature in the observed distribution of galaxies, mostly due to the large-scale bulk flows \citep[e.g.][]{2007ApJ...664..660E}, degrading the accuracy of measured BAO scale.
BAO reconstruction has been proposed by \citet{2007ApJ...664..675E} to reduce the nonlinear degradation effects by reversing the large-scale shifts and to recover the linear BAO signal.
The standard BAO reconstruction method has been tested in simulations \citep[e.g.][]{2007ApJ...664..675E,2010ApJ...720.1650S,2011ApJ...734...94M,2012JCAP...10..006T,2014MNRAS.445.3152B,2015PhRvD..92l3522S}, explored for modeling with perturbation theory \citep[e.g.][]{2009PhRvD..79f3523P,2009PhRvD..80l3501N,2015MNRAS.450.3822W,2016MNRAS.460.2453S,2017PhRvD..96d3513H,2019arXiv190700043C} 
and applied in galaxy clustering analysis, such as SDSS \citep{2012MNRAS.427.2132P}, BOSS \citep[][]{2012MNRAS.427.3435A,2014MNRAS.441...24A,2014MNRAS.440.2222T,2016MNRAS.455.3230B,2017MNRAS.470.2617A}, WiggleZ \citep{2014MNRAS.441.3524K,2016MNRAS.455.3230B},
SDSS MGS \citep{2015MNRAS.449..835R}, and 6dFGS \citep{2018MNRAS.481.2371C}.
Recently, several nonlinear reconstruction algorithms have been developed to improve upon the standard method \citep{2017PhRvD..96l3502Z,2017PhRvD..96b3505S,2018PhRvD..97b3505S,2018MNRAS.478.1866H} and have been applied to simulated halo/galaxy fields \citep{2017ApJ...847..110Y,2019ApJ...870..116W,2019MNRAS.483.5267B,2019MNRAS.482.5685H}.
Forward modeling methods have also been explored \citep{2017JCAP...12..009S,2018JCAP...10..028M,2019JCAP...01..042S,2019arXiv190607143E,2019arXiv190702330M}.

The key ingredient for reconstruction is an estimate of the displacement field from the observed galaxy density field. 
In standard reconstruction, the estimated displacement field is used to move galaxies to the initial positions, restoring the linear BAO signal.
For most nonlinear reconstruction algorithms, we directly use the reconstructed nonlinear displacement to measure the BAO signal \citep[e.g.][]{2017PhRvD..96l3502Z,2017PhRvD..96b3505S,2018PhRvD..97b3505S}.
The conventional way to estimate the displacement field is first to embed the survey volume in a periodic box, then smooth the nonlinear density field using a Gaussian window with scale $\sim\,10\,h^{-1}\mr{Mpc}$ to suppress small-scale nonlinearities. 
The large-scale displacement field is obtained by solving the linearized continuity equation relating density and displacement \citep[e.g.][]{2012MNRAS.427.2132P}.
This simple displacement estimation method has been shown to be close to optimal for galaxy surveys with low number densities and large survey volumes like the SDSS BOSS survey \citep[e.g.][]{2018MNRAS.477.1153V}.
However, the linear operator involved is an inverse power of the wavenumber, $\sim k^{-2}$, which makes the reconstruction of the displacement field from the observed density a nontrivial process near the survey boundary.

The next generation surveys such as DESI-BGS will have much lower shot noises than the current data.
However, for these dense low redshift galaxy surveys with relative smaller volumes, the boundary effect will become important for a large fraction of the total volume and improvements near the boundary would be highly beneficial if external data can be used for estimating the displacement.

The peculiar velocities of observed galaxies also provides a way to infer the displacement field using the same Wiener filtering process as reconstruction with density fields \citep[e.g.][]{1999ApJ...520..413Z,2013MNRAS.430..888D}.
While the displacement-density relation is nontrivial (even in the linear regime) due to the inverse Laplacian $k^{-2}$, the displacement-velocity relation is a simple scaling (with $afH$) in linear theory. 
Without measurement errors, and in linear theory, we can have perfect reconstruction of the displacements near the survey boundary as long as we know the velocities at the same position.
In the presence of observational noise, the large-scale correlations of peculiar velocities still allow us to infer the reconstructed fields from the velocities far from the boundary.
The kinematic Sunyaev-Zeldovich (kSZ) effect \citep{1972CoASP...4..173S,1980MNRAS.190..413S,1987ApJ...322..597V} offers a unique opportunity to measure peculiar velocities at cosmological distances \citep[e.g.][for recent investigations]{2009arXiv0903.2845H,2011MNRAS.413..628S}.
Several measurements have been made with Planck \citep[e.g.][]{2016A&A...586A.140P,2015PhRvL.115s1301H,2017arXiv171208619L,2018PhRvD..97b3514L,2016PhRvL.117e1301H,2016PhRvD..94l3526F}
and ACT data \citep[e.g.][]{2012PhRvL.109d1101H,2016PhRvD..93h2002S,2017JCAP...03..008D}.

By combining future optical surveys (such as DESI, \citealt{2016arXiv161100036D}, Euclid, \citealt{2018LRR....21....2A} and LSST, \citealt{2009arXiv0912.0201L}) and Cosmic Microwave Background (CMB) experiments (such as the future Simons Observatory, \citealt{2019JCAP...02..056A}, and CMB-S4, \citealt{2016arXiv161002743A}) the radial peculiar velocities can be measured with high precision from the kSZ effect imprinted on the CMB for millions of galaxies and can be used to constrain cosmological parameters \citep[e.g.][]{2018arXiv181013423S, 2018arXiv181013424M, 2018PhRvD..98l3501D,2018PhRvD..98f3502C,2019arXiv190706678M,2019arXiv190604208P}.

In this paper, we study the effect of the survey boundary on the estimation of the displacement from the observed density field and explore the improvement with the inclusion of velocities from the kSZ effect measurement. 
We find that the velocity information can help both low redshift dense surveys, where the volume is small, and high redshift, low number density surveys where the estimated displacement is usually noisy.

In Section~\ref{sec:formalism}, we present the formalism of Wiener filtering for reconstructing displacements from density and velocity fields.
Section~\ref{sec:1d} provides a one-dimensional toy model to illustrate the boundary effect and the idea of combining density and velocity to estimate the displacement. 
In Section~\ref{sec:3d}, we show the results for the three-dimensional case and estimate the improvements with future large-scale structure and CMB surveys.
We discuss the future prospects and conclude in Section~\ref{sec:cls}.

\section{Wiener filter reconstruction}
\label{sec:formalism}

The reconstruction of the displacement field can be implemented with the Wiener filtering formalism \citep[e.g.][]{1992ApJ...398..169R,1995ApJ...449..446Z,1995MNRAS.272..885F} and has been applied in the galaxy clustering analysis of the BAO signal with standard reconstruction by \citet{2012MNRAS.427.2132P}, where a constrained realization is used to fill in unobserved regions.
However, for most large volume low number density surveys such SDSS-BOSS, since the field is noisy and the boundary effect is small, we usually opt for a simpler and faster method works as follows:
\begin{enumerate}
\item Embed the survey volume in a larger box and fill the unobserved region with uniform density distribution.
\item Smooth the density with a Gaussian window function of scale $\sim10\,h^{-1}\mr{Mpc}$ to suppress the small-scale nonlinearities and shot noises.
\item Compute the displacement field from the smoothed density field using the linear continuity equation.
\end{enumerate}
This simple method is very fast and is close to optimal for large volume surveys with low number densities.
However, the drawback of this method is that it does not separate the signal, the part of density field correlated with the displacement field, and the noise including shot noise and small-scale stochastic nonlinearities, which are not correlated with the displacement.  The method relies on the Gaussian smoothing to mitigate both the effects of shot noise and nonlinearities. 
Also, using the same Gaussian smoothing for the whole box assumes the noise is constant well-inside and outside the observed region.
In reality, we can think of the noise as being small in the deep interior of the survey and infinite in the padding region when we express the problem in a larger enclosing box.
Therefore, the noise properties become highly anisotropic near the survey boundary, even if the observed galaxy density is constant in the observed region.
In the limit of an arbitrarily large survey volume the boundary effect is small and isotropic smoothing suffices. 

The Wiener filtering approach provides a way to do dynamical reconstruction, i.e., using the observed data which noisily samples one field (such as the galaxy density field) to reconstruct another field (such as the displacement or velocity field) which is dynamically related \citep[see e.g.][]{1995ApJ...449..446Z}.
In this section, we present the formalism of Wiener filtering and its application to displacement estimation from density and velocity fields.  Our presentation will largely follow \citet{1995ApJ...449..446Z}.

We consider a set of observations, $\bmd=\{d_i\}$ ($i=1,\dots,M$) sampled at positions $\bmx_i$, which measure an underlying field, $\bms=\{s_\alpha\}$ ($\alpha=1,\dots,N$),
\begin{equation}
    \bmd=\bmR\bms+\bmeps,
\end{equation}
where $R$ is an $M\times N$ matrix which represents a response function and $\bmeps=\{\epsilon_i\}$ ($i=1,\dots,M$) are the statistical errors of the observations.
Assuming such a linear model we can write the minimum variance, unbiased linear estimator of $\bm{s}$, i.e.~$\bms^{\mr{MV}}$, as 
\begin{equation}
    \bms^{\mr{MV}}=\bmF\bmd,
\end{equation}
where $F$ is an $N\times M$ matrix.
The filter $\bmF$ can be found by minimizing the variance of the residual vector $\bmr$,
\begin{equation}
    \langle\bmr\bmr^{\dag}\rangle=\langle(\bms-\bms^{\mr{MV}})(\bms^{\dagger}-\bms^{\mr{MV}\dagger})\rangle,
\end{equation}
under the constraint that $\langle\bm{s}^{\rm MV}\rangle=\bm{s}$, to obtain
\begin{equation}
    \bmF=\langle\bms\bmd^\dagger\rangle\langle\bmd\bmd^\dagger\rangle^{-1},
\end{equation}
which is usually referred as the Wiener filter.
The variance of the residual of the $\alpha$th degree of freedom of the underlying field is 
\begin{equation}
    \langle|r_\alpha|^2\rangle=\langle|s_\alpha|^2\rangle-\langle s_\alpha\bmd^\dagger\rangle\langle\bmd\bmd^\dagger\rangle^{-1}\langle\bmd s_\alpha^*\rangle,
\end{equation}
which describes the error of the estimated field $s_\alpha$ at position $\bmx_\alpha$.
Note that the position $\bmx_\alpha$ can be a measured point within the survey volume as well as an unobserved point outside the survey volume.
The Wiener filtering allows us to extrapolate the reconstructed field into a larger domain not covered by observed galaxies.
The Wiener filter approach is linear estimation based on the principle of minimum mean squared error and for Gaussian random fields it coincides with the maximum posterior Bayesian estimation of the underlying field.
Here we only summarize the ingredients of Wiener filtering used in this paper.
For a more detailed description of the Wiener filtering approach, see \citet{1995ApJ...449..446Z}.

In our case, the underlying field $\bms$ to be reconstructed is the linear displacement field $\bmp$, with the observed density field $\hat{\delta}$ and velocity field $\hat{\bmv}$.
The operational procedures of underlying field reconstruction are first inversion of the data covariance $\langle\bmd\bmd^\dagger\rangle$ and then multiplication of the cross-correlation function $\langle\bms\bmd^\dagger\rangle$.
For estimating the displacement field from the density field, we have
\begin{equation}
    \bmp^{\mr{WF}}(\bmx)=\langle\bmp(\bmx)\delta(\bmx_i)\rangle\langle\hat{\delta}(\bmx_i)\hat{\delta}(\bmx_j)\rangle^{-1}\hat{\delta}(\bmx_j).
\end{equation}
Note that the displacement $\bmp(\bmx)$ is a vector field.
In the one-dimensional case, a vector has one component and in the three-dimensional space it has three components ($x,y,z$).
In the following discussions, we use the indices $\mu,\nu=x,y,z$ to denote one component of the vector fields like the displacement and velocity.
The variance of the residual for displacement in the $\mu$ direction $\psi_\mu(\bmx)$ at position $\bmx$ is given by 
\begin{equation}
    \langle\Delta\psi_\mu^2(\bmx)\rangle=\langle\psi_\mu^2\rangle-\langle \psi_\mu(\bmx)\delta(\bmx_i)\rangle\langle\hat{\delta}(\bmx_i)\hat{\delta}(\bmx_j)\rangle^{-1}\langle\delta(\bmx_j) \psi_\mu(\bmx)\rangle,
\end{equation}
where $\Delta\psi_\mu(\bmx)=\psi_\mu(\bmx)-\psi_\mu^{\mr{WF}}(\bmx)$ is the residual of the reconstructed displacement.
We have assumed the noise term $\bmeps$ is not correlated with the signal and thus does not contribute to the cross-correlation matrices.
Note that the variance of the residual for $\psi_\mu(\bmx)$ depends on its position $\bmx$, mostly the distance to the survey boundary.

When we instead use the velocity field for reconstruction, we have
\begin{equation}
    \bmp^{\mr{WF}}(\bmx)=\langle\bmp(\bmx)v_\nu(\bmx_i)\rangle\langle\hat{v}_\nu(\bmx_i)\hat{v}_\nu(\bmx_j)\rangle^{-1}\hat{v}_\nu(\bmx_j).
\end{equation}
Similarly, the uncertainties of the reconstructed displacement $\psi_\mu(\bmx)$ is 
\begin{equation}
    \langle\Delta\psi_\mu^2(\bmx)\rangle=\langle\psi_\mu^2\rangle-
    \langle \psi_\mu(\bmx)v_\nu(\bmx_i)\rangle\langle\hat{v}_\nu(\bmx_i)\hat{v}_\nu(\bmx_j)\rangle^{-1}\langle v_\nu(\bmx_j) \psi_\mu(\bmx)\rangle,
\end{equation}
where the $\psi_\mu$ and $v_\nu$ can be in the same direction or different directions.
When $\mu=\nu$, we are considering the reconstruction of displacement in the same direction as the observed velocity.
Therefore, from the observed velocities, e.g.\ the radial velocities from the kSZ effect measurement, we can infer the the displacement in the line of sight direction.
However, the correlation between the displacement and velocity in different directions $\mu\neq\nu$ still allows us to reconstruct the transverse displacements which are perpendicular to the line-of-sight direction from the radial velocity, though with larger errors due to the weaker correlation.

We can also combine the observations of the density and velocity fields for reconstructing the displacement field as
\begin{equation}
    \bmp^\mr{WF}=\left( \begin{array}{cc}
\langle\bmp\delta\rangle & \langle\bmp v_\mu\rangle
\end{array} \right)\hat{C}^{-1}\left( \begin{array}{c}
\hat{\delta}  \\
\hat{v}_\mu
\end{array} \right),
\end{equation}
where the $2M\times2M$ covariance matrix is
\begin{equation}
    \hat{C}=\left( \begin{array}{cc}
\langle\hat{\delta}\hat{\delta}\rangle & \langle\delta v_\mu\rangle \\
\langle v_\mu\delta\rangle & \langle\hat{v}_\mu\hat{v}_\mu\rangle
\end{array} \right).
\end{equation}
The uncertainties of the reconstructed field is
\begin{eqnarray}
\label{eq:res_vd}
\langle\Delta\psi_\mu^2\rangle=\langle\psi_\mu^2\rangle-
\left( \begin{array}{cc}
\langle\psi_\mu\delta\rangle & \langle\psi_\mu v_\nu\rangle
\end{array} \right)\hat{C}^{-1}\left( \begin{array}{c}
\langle\delta\psi_\mu{}\rangle  \\
\langle v_\nu\psi_\mu\rangle
\end{array} \right).
\end{eqnarray}
We expect that the improvement from including the velocity field will depend on the relative noise of the measured density and velocity fields and the distance to the boundary.
To explore the boundary effects on the displacement estimation, the above equations have to be computed in configuration space instead of Fourier space.  While the computation in Fourier space is fast, it requires homogeneity and isotropy of the underlying field and shot noise which are not satisfied with the real observed galaxy density especially for small volume surveys.

In the next section, we begin with a toy model in one-dimension to illustrate the effect of survey boundary on the reconstruction and the idea of combining density and velocity.
The dimensionality of the one-dimensional problem allows us to invert the covariance matrix directly using Cholesky decomposition.  In three dimensions we need to invert a matrix with dimensions $M\sim\mathcal{O}(10^7)$.
We use the preconditioned conjugate gradient method to perform the inverse \citep[see e.g.][]{shewchuk1994introduction,2003NewA....8..581P,2012MNRAS.427.2132P}.

\section{The 1D toy model}
\label{sec:1d}

The dynamical reconstruction including the reconstruction of the underlying density field from the observed radial peculiar velocities or using the observed density field to construct the displacement and velocity fields relies on a theoretical model which relates the two different fields.
The continuity equation describes the relation between velocity and density.
The linearized continuity equation is given by
\begin{equation}
\dot{\delta}+\nabla\cdot\bm{v}=0,
\end{equation}
where the dot denotes partial derivative with respect to conformal time.
In Fourier space, we have 
\begin{equation}
\bm{v}(\bm{k})=\frac{i\bm{k}afH}{k^2}\delta(\bmk),
\end{equation}
where we have assumed potential flow and linear perturbation theory.
The linear displacement under the Zel'dovich approximation is given by 
\begin{equation}
    \delta+\nabla\cdot\bmp=0,
\end{equation}
and in Fourier space we have
\begin{equation}
    \bmp(\bm{k})=\frac{i\bm{k}}{k^2}\delta(\bmk).
\end{equation}
The linear operator involved above is an inverse power of the wavenumber $k^{-2}$ and can be thought as an integration of the density field over the surrounding region.
This makes the displacement-from-density reconstruction depend non-locally on the observed density field, while the velocity-to-density process involves only a differential operator which is local and therefore only depends on the local values of the observed velocity field. 
However, the velocity-displacement relation is quite simple in linear theory,
\begin{equation}
    \bmv=afH\bm{\psi},
\end{equation}
where the two quantities are related by a coefficient which depends on the fiducial background cosmology.

In one dimension, the displacement only has one component $\psi$.
The lower dimensionality allows us to get an intuitive picture of the problem \citep{2016JCAP...01..043M}.
In the following discussion, we consider the reconstruction problem in one dimension and assume the reflection symmetry for the power spectrum in the 1D space, i.e., $P_\mr{1D}(k)=P_\mr{1D}(-k)$.
We consider the displacement power spectrum
\begin{equation}
    P_\psi(k)=A\exp(-k^2R^2/2),
\end{equation}
where $A$ is the normalization factor and $R=50\,\mr{Mpc}/h$ is the correlation length of the displacement, roughly corresponding to the correlation length of the velocity/displacement fields in three dimensions in a cosmology with up-to-date cosmological parameters.
The variance of the displacement in the 1D is given by
\begin{equation}
    \langle\psi(x)\psi(x)\rangle=\int_0^\infty\frac{dk}{\pi}P_\psi(k).
\end{equation}
We normalize the power spectrum by choosing the variance of the displacement equal to the 3D case, i.e.,
\begin{equation}
    \frac{A}{\sqrt{2\pi}R}\simeq100\ (\mr{Mpc}/h)^2
    \quad\Rightarrow\quad
    A\simeq1.2530\times10^4\ (\mr{Mpc}/h)^3.
\end{equation}
The corresponding density power spectrum is given by
\begin{equation}
\label{eq:pk_d_1d}
    P_\delta(k)=k^2P_\psi(k).
\end{equation}
and note that the observed density power spectrum also includes shot noise.
In the 1D case, the correlation function can be computed by
\begin{equation}
\xi(r)=\int_{-\infty}^{+\infty}\frac{dk}{2\pi}P_{\mr{1D}}(k)\exp{(ikr)},
\end{equation}
where $P_{\mr{1D}}(k)$ is the power spectrum in the 1D space.
{
The displacement correlation function is given by
\begin{equation}
    \langle\psi(x_1)\psi(x_2)\rangle=\frac{A}{\sqrt{2\pi}R}\exp\bigg[\frac{-(x_1-x_2)^2}{2R^2}\bigg].
\end{equation}
The density-displacement cross correlation function is 
\begin{equation}
    \langle\delta(x_1)\psi(x_2)\rangle=\frac{A}{\sqrt{2\pi}R}\exp\bigg[\frac{-(x_1-x_2)^2}{2R^2}\bigg]\frac{x_1-x_2}{R^2},
\end{equation}
and we also have the displacement-density cross-correlation function
\begin{equation}
    \langle\psi(x_1)\delta(x_2)\rangle=\frac{A}{\sqrt{2\pi}R}\exp\bigg[\frac{-(x_1-x_2)^2}{2R^2}\bigg]\frac{x_2-x_1}{R^2},   
\end{equation}
Finally, the density correlation function is 
\begin{equation} \label{eq:delta_1d}
    \langle\delta(x_1)\delta(x_2)\rangle=\frac{A}{\sqrt{2\pi}R}\exp\bigg[\frac{-(x_1-x_2)^2}{2R^2}\bigg]\frac{1-(x_1-x_2)^2/R^2}{R^2}.
\end{equation}
}
The data covariance matrix includes both the density correlation and shot noise,
\begin{equation}
\label{eq:cov_d_1d}
    \langle\hat{\delta}(x_i)\hat{\delta}(x_j)\rangle=\langle\delta(x_i)\delta(x_j)\rangle+\frac{\delta^D(x_i-x_j)}{\bar{n}(x_i)},
\end{equation}
where $\delta^D(x)$ is the Dirac delta function.

For reconstruction with densities, we have the residual variance of the displacement
\begin{equation}
\label{eq:res_d_1d}
    \langle\Delta\psi^2(r)\rangle=\langle\psi^2(r)\rangle-\langle\psi(r){\delta}(r_i)\rangle\langle\hat{\delta}(r_i)\hat{\delta}(r_j)\rangle^{-1}\langle{\delta}(r_j)\psi(r)\rangle.
\end{equation}
Consider the ideal case where the survey boundary does not affect the reconstruction, for example in a region well inside the survey volume or using data from $N$-body simulations with periodic boundary conditions.  The residual variance should not depend on the position vector $r$.
In the case of periodic boundary conditions, we can transform the fields to Fourier space.
The data covariance for the density field is simply $P_\delta(k)+1/\bar{n}$ and the density-displacement cross-correlation is just the cross power spectrum, $P_{\delta\psi}(k)$.
We have 
       \begin{equation}
    \langle\hat{\delta}(r_i)\hat{\delta}(r_j)\rangle^{-1}=\int\frac{dk}{2\pi} \frac{1}{(P_\delta(k) + 1/\bar{n}}\exp[ik(r_i-r_j)],
\end{equation}
and $\langle\hat{\delta}(r_i)\hat{\delta}(r_k)\rangle\langle\hat{\delta}(r_k)\hat{\delta}(r_j)\rangle^{-1}=I_{ij}$, where $I$ is the identity matrix.
In Fourier space we have the relation
\begin{equation}
\label{eq:res_d_1d_k}
    \langle\Delta\psi^2(r)\rangle=\int_0^\infty\frac{dk}{\pi}P_\psi(k)-\int_0^\infty\frac{dk}{\pi}P_\psi(k)\frac{P_\delta(k)}{P_\delta(k)+1/\bar{n}},
\end{equation}
where we have assumed the density and displacement are fully correlated.
When the number density is infinite, i.e., the measurement noise vanishes ($1/\bar{n}=0$) the residual variance is zero, which means we have a perfect measurement of the displacement field.
Note that this is under the assumption of linear theory.
When the measurement noise is nonzero, we will have a nonzero reconstruction error.
With higher number density, and thus lower $1/\bar{n}$, we have a better reconstruction of the displacement field.

However, for a general point $r$, we can not simplify the covariance matrix inversion of equation~(\ref{eq:res_d_1d}) as an integral in equation~(\ref{eq:res_d_1d_k}) since the covariance is no longer diagonal.
We consider the density field observed in a 1D box of length $L$ and the densities outside are not measured.
For a point at $r=L/2$, the reconstruction error is given by equation~(\ref{eq:res_d_1d_k}) for large $L$.
When $r=0$, we can only infer the displacement from the densities on the right hand of this point, while for $r=L/2$ we can use the density data from both sides.
For reconstruction with lower noise, we expect that the residual variance is roughly twice as large at the boundary $r=0$ or $L$ compared to $r=L/2$,
\begin{equation}
    \langle\Delta\psi^2(r)\rangle\approx2\int_0^\infty\frac{dk}{\pi}P_\psi(k)\frac{1/\bar{n}}{P_\delta(k)+1/\bar{n}},
\end{equation}
since we only have half the data points to reconstruct the displacement.
When the noise is very large, the residual variance is dominated by the cosmic variance everywhere, both inside and outside the box, thus we are not reconstructing anything from the observation.

\begin{figure}
	\includegraphics[width=\columnwidth]{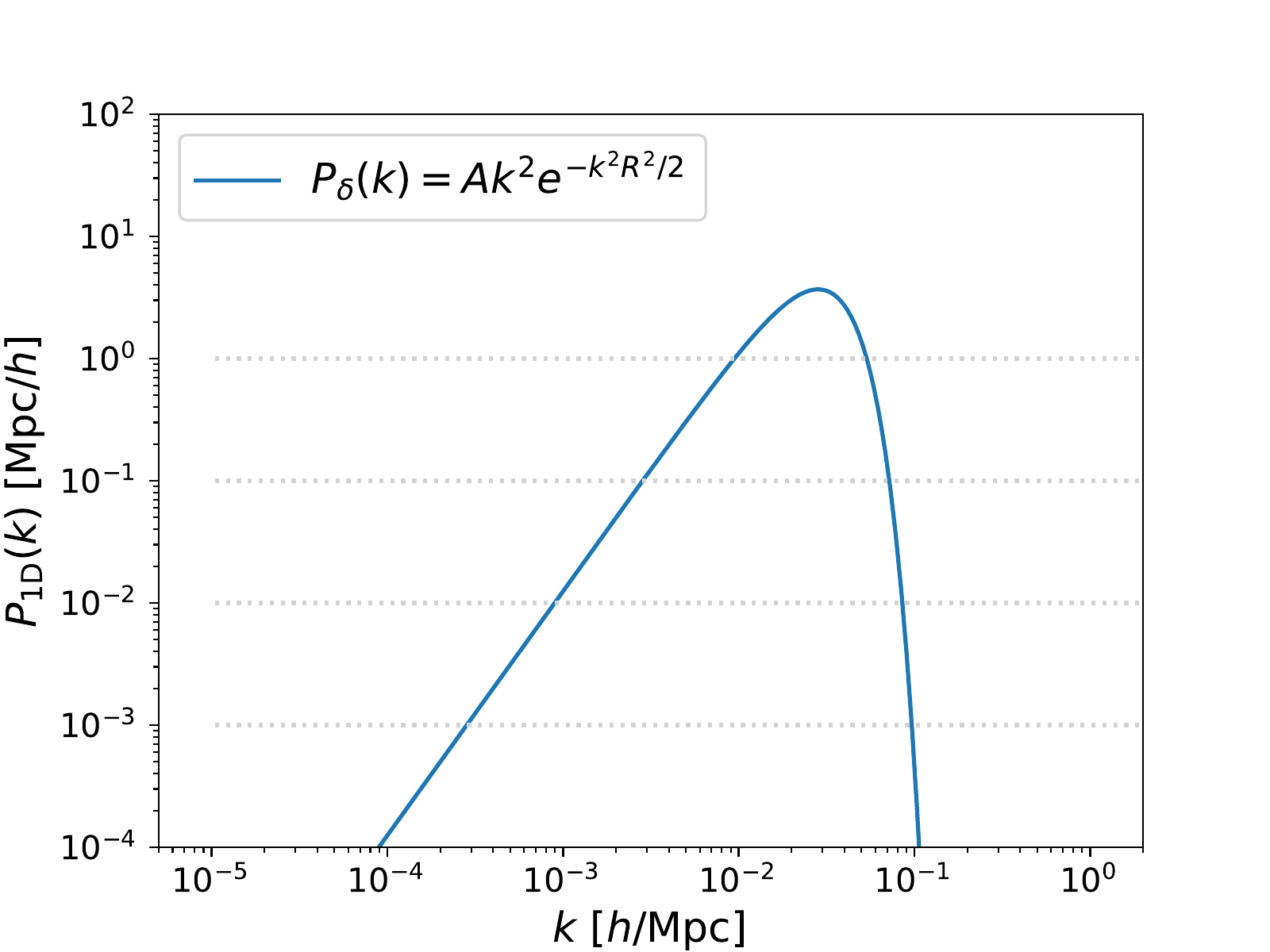}
    \caption{The density power spectrum in 1D space and the shot noise power, $1/\bar{n}$, for $\bar{n}=1$, 10, $10^2$, $10^3\,h\,\mathrm{Mpc}^{-1}$, respectively.}
    \label{fig:pk_d_1d}
\end{figure}

\begin{figure}
	\includegraphics[width=\columnwidth]{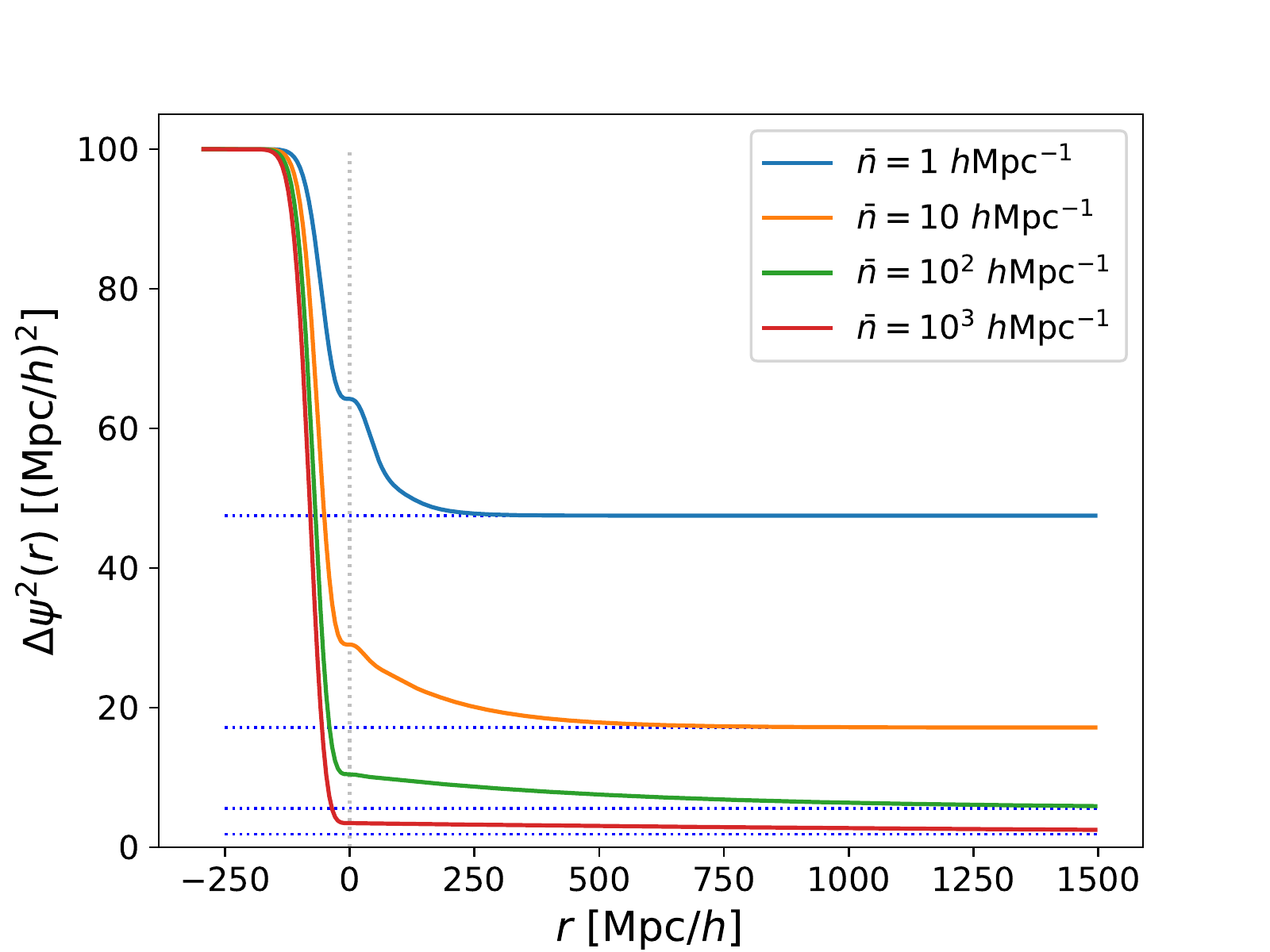}
    \caption{The residual variance $\langle\Delta\psi^2(r)\rangle$ for the reconstructed displacement with different density noises in the 1D space. The dotted lines show the ideal cases where residual variance only arises from the shot noise. In the low noise limit, the variance at $r=0$ is roughly twice compared to the ideal case.}
    \label{fig:v_pd_1d}
\end{figure}

In the numerical calculation, we take the box size $L=1.2\times10^4\,\mr{Mpc}/h$ sampled on a regular grid of $N=2048$ points.
We have tested that the results converge with this configuration and the points inside the 1D box converge to the variance computed in Fourier space given by equation~(\ref{eq:res_d_1d_k}).
In Figure \ref{fig:pk_d_1d}, we show the density power spectrum given in equation~(\ref{eq:pk_d_1d}) and shot noise power for $\bar{n}=1$, 10, $10^2$, $10^3\,h\,\mr{Mpc}^{-1}$, respectively.
The observed density covariance can be directly computed by equation~(\ref{eq:cov_d_1d}), with the density correlation function given by Eq.~(\ref{eq:delta_1d}).
The inversion of a covariance matrix of size $N^2\sim10^6$ can be achieved through Cholesky decomposition. 
In Figure \ref{fig:v_pd_1d}, we plot the residual variance $\langle\Delta\psi^2(r)\rangle$ of the displacement field from reconstruction as a function of the distance to the boundary for different shot noises.
The dashed line shows the residual variance computed in Fourier space by equation~(\ref{eq:res_d_1d_k}), where the reconstruction error only arises from the measurement noise instead of the boundary effect. 
We see that for points within the survey, i.e., $r\gg0$, the variance converges to the Fourier space result and does not depend on the position $r$.
For a point near the survey boundary, $r\sim0$, the underlying field and noise are no longer homogeneous or isotropic. 
The residual variance calculated by equation~(\ref{eq:res_d_1d}) depends on the distance to the boundary and is quite different from equation~(\ref{eq:res_d_1d_k}).
We see that the residual variance is about twice as large at the boundary for the lower noise cases as expected.
When $r\ll0$, the residual variance is basically the cosmic variance as we have no data to infer the displacement at those positions.

If we can also measure the velocities at the same positions where we have measured the densities, we can use the velocity information to reconstruct the displacement field.
In linear theory, the two fields are related by the factor $afH$ and the velocity power spectrum is 
\begin{equation}
    P_v(k)=(afH)^2P_\psi(k).
\end{equation}
We have the velocity-displacement correlation function
\begin{equation}
    \langle v(x)\psi(x+r)\rangle=afH\frac{A}{\sqrt{2\pi}R}\exp\bigg(\frac{-r^2}{2R^2}\bigg),
\end{equation}
and the velocity correlation function is given by
\begin{equation}
    \langle v(x)v(x+r)\rangle=(afH)^2\frac{A}{\sqrt{2\pi}R}\exp\bigg(\frac{-r^2}{2R^2}\bigg).
\end{equation}
The observed velocity covariance matrix includes both the signal and the measurement noise,
\begin{equation}
    \langle\hat{v}(x_i)\hat{v}(x_j)\rangle=\langle v(x_i)v(x_j)\rangle+N_v(x_i)\delta^D(x_i-x_j),
\end{equation}
where $N_v(x_i)=\langle\epsilon_i^2\rangle$ is the variance of the galaxy velocity errors.
Here, we assume the errors of different data points are statistically independent.
Similarly, for reconstruction with the velocities, we have the residual variance 
\begin{equation}
\label{eq:res_v_1d}
    \langle\Delta\psi^2(x)\rangle=\langle\psi^2\rangle-
    \langle \psi(x)v(x_i)\rangle\langle\hat{v}(x_i)\hat{v}(x_j)\rangle^{-1}\langle v(x_j) \psi(x)\rangle.
\end{equation}
When the covariance is diagonal in Fourier space, we have
\begin{equation}
\label{eq:res_v_1d_k}
    \langle\Delta\psi^2(r)\rangle=\int_0^\infty\frac{dk}{\pi}P_\psi(k)-\int_0^\infty\frac{dk}{\pi}P_\psi(k)\frac{P_v(k)}{P_v(k)+N_v},
\end{equation}
where we have assumed the velocity and displacement are fully correlated.

\begin{figure}
	\includegraphics[width=\columnwidth]{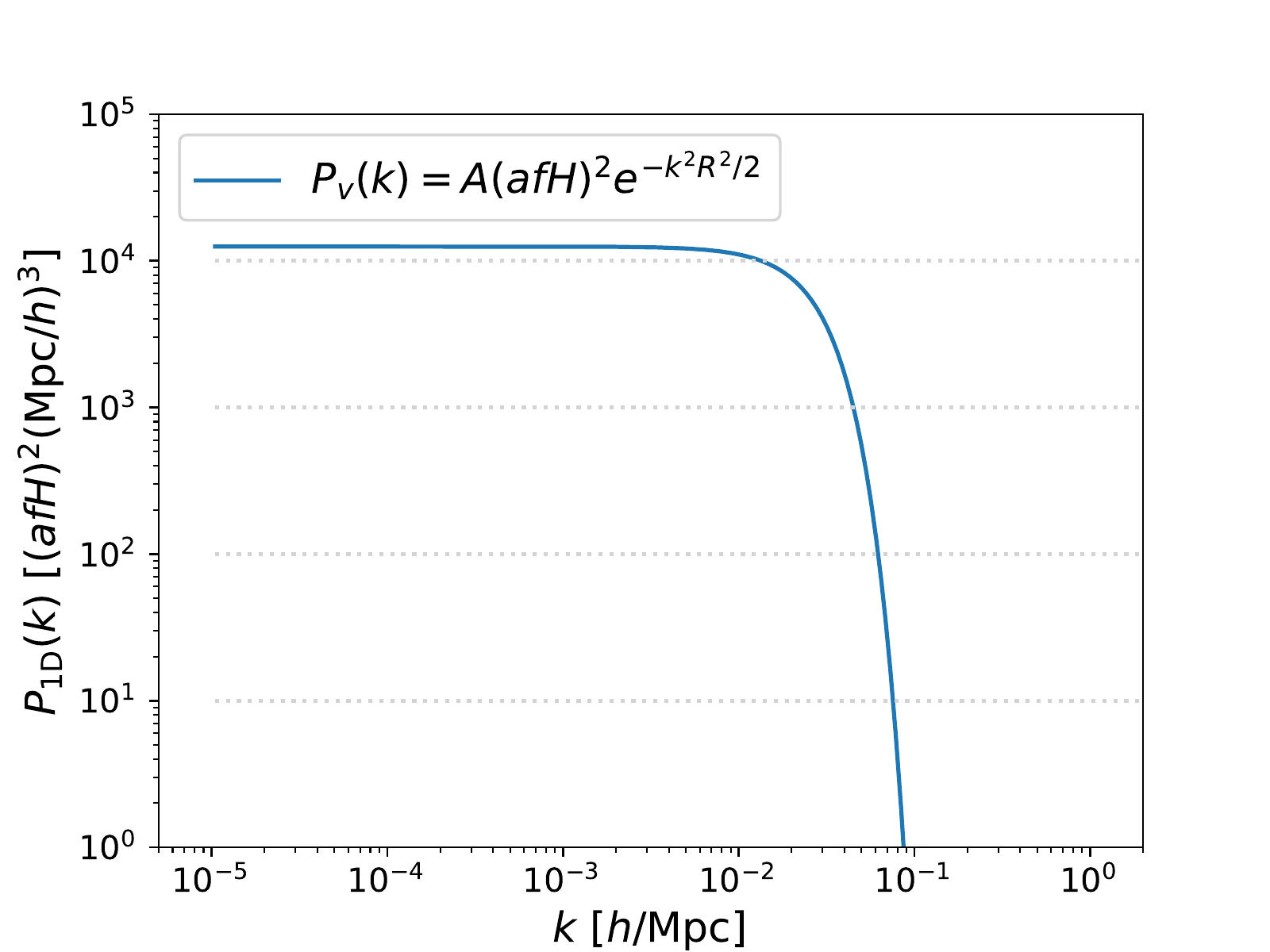}
    \caption{The velocity power spectrum for the 1D toy model and velocity noise level $N_v/(afH)^2=10$, $10^2$, $10^3$, $10^4\ (\mathrm{Mpc}/h)^3$.}
    \label{fig:pk_p_1d}
\end{figure}

\begin{figure}
	\includegraphics[width=\columnwidth]{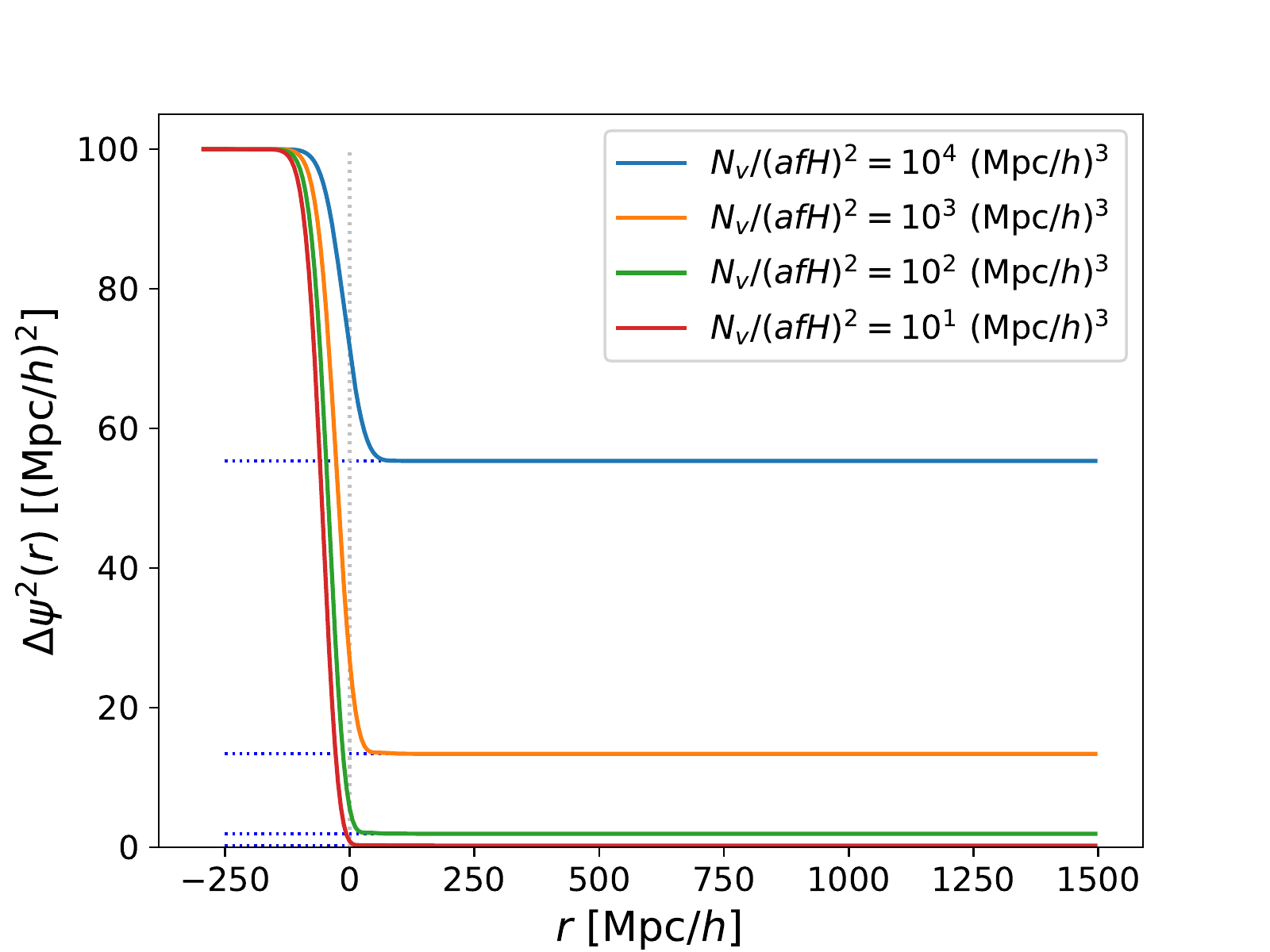}
    \caption{The residual variance $\langle\Delta\psi^2(r)\rangle$ for the reconstructed displacement with different velocity noises in the 1D space. The dotted lines show the ideal case where residual variance only arises from the observational errors. In contrast to reconstruction from densities the variance is not affected much by the boundary.}
    \label{fig:v_pp_1d}
\end{figure}

Figure \ref{fig:pk_p_1d} shows the velocity power spectrum and noise power spectra $N_v/(afH)^2=10$, $10^2$, $10^3$, $10^4\,(\mr{Mpc}/h)^3$.
In Fig.~\ref{fig:v_pp_1d}, we show the residual displacement errors for reconstruction from velocities as a function of the distance to the boundary.
We find that for reconstruction with smaller noise levels, the residual variance is almost unaffected by the boundary.
This is because of the linear relation between displacement and velocity; a noiseless measurement of the velocity at $r$ can give a perfect estimation of the displacement at the same position.
For higher noise cases the residual variance becomes larger near the boundary but the scales affected are still much smaller than for the densities.

We can combine the density and velocity measurements for estimating the displacement field.
The uncertainties of the reconstructed field are given by
\begin{eqnarray}
\langle\Delta\psi_\mu^2\rangle=\langle\psi_\mu^2\rangle-
\left( \begin{array}{cc}
\langle\psi\delta\rangle & \langle\psi v\rangle
\end{array} \right)\hat{C}^{-1}\left( \begin{array}{c}
\langle\delta\psi{}\rangle  \\
\langle v\psi\rangle
\end{array} \right),
\end{eqnarray}
where the covariance matrix
\begin{equation}
    \hat{C}=\left( \begin{array}{cc}
\langle\hat{\delta}\hat{\delta}\rangle & \langle\delta v\rangle \\
\langle v\delta\rangle & \langle\hat{v}\hat{v}\rangle
\end{array} \right).
\end{equation}
When the blocks of the covariance are diagonal in Fourier space, the covariance matrix can be inverted blockwise and we have 
\begin{eqnarray}
\label{eq:res_comb_1d_k}
    \langle\Delta\psi^2(r)\rangle=\int_0^\infty\frac{dk}{\pi} P_\psi(k)-\int_0^\infty\frac{dk}{\pi}\frac{P_\psi P_\delta}{P_{\hat{\delta}}-P_\delta P_v/P_{\hat{v}}}\nonumber\\
    -\frac{2P_\psi P_vP_\delta}{P_{\hat{\delta}}(P_{\hat{v}}-P_vP_\delta/P_{\hat{\delta}})}+\frac{P_\psi P_v}{P_{\hat{v}}-P_vP_\delta/P_{\hat{\delta}}},
\end{eqnarray}
where $P_{\hat{\delta}}=P_\delta+1/\bar{n}$ and $P_{\hat{v}}=P_v+N_v$.
When the velocity noise is infinite, $N_v\to\infty$, the above equation reduces to equation~(\ref{eq:res_d_1d_k}), i.e. the velocity does not help with the reconstruction.
The improvement by including the velocity information depends on the relative noise level of the measured density and velocity fields.

\begin{figure}
	\includegraphics[width=\columnwidth]{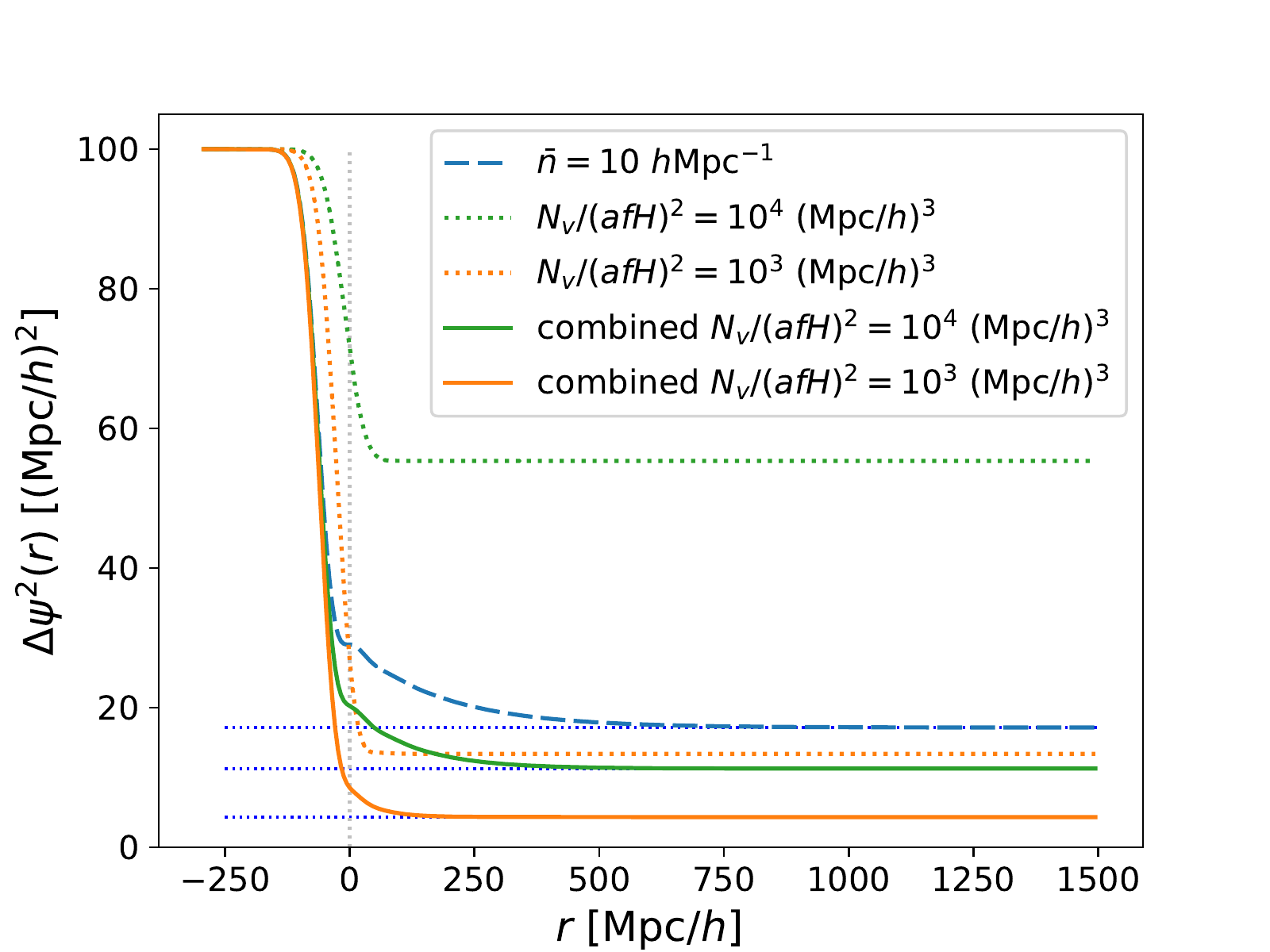}
    \caption{The residual variance $\langle\Delta\psi^2(r)\rangle$ for the reconstructed displacement with densities and velocities. Even when the velocity estimate is noisy, combining with the velocity measurement can still improve the performance. As the velocity measurement improves we can reduce the residual error near the boundary significantly.
    }
    \label{fig:v_combine}
\end{figure}

Fig.~\ref{fig:v_combine} shows the results for reconstruction with both the density and velocity fields.
The dotted horizontal lines show the results computed using equation~(\ref{eq:res_comb_1d_k}) and Table~\ref{tab:res_k} summarizes the numbers. 
We find that even a noisy velocity measurement can still improve the estimation of the displacement field, reducing the variance by $\sim30\%$.
The residual variance at the boundary is also reduced to the value inside the box when we use only the densities.
When we have a better measurement of the velocity, comparable to the density field in the sense of the similar displacement residual variance, the estimation of the displacement field can be improved significantly. 
We notice that the residual variance with both velocity and density fields is better than a naive inverse sum of two independent pieces of information because of the correlation between density and velocity fields.

\begin{table}
	\centering
	\caption{The residual variance computed by equation~(\ref{eq:res_comb_1d_k}) for different density and velocity noises, with the unit $(\mr{Mpc}/h)^2$.}
	\label{tab:res_k}
	\begin{tabular}{cccc} 
		\hline
		$N_v/(afH)^2\,[(\mr{Mpc}/h)^3]$ & $10^3$ & $10^4$ & $\infty$ \\
		\hline
		$\bar{n}=10\,h\mr{Mpc}^{-1}$ & 4.33 & 11.27 & 17.15\\
		$\bar{n}=0\,h\mr{Mpc}^{-1}$ & 13.38 & 55.34 & 100\\
		\hline
	\end{tabular}
\end{table}

\section{The three-dimensional situation}
\label{sec:3d}

\begin{figure}
	\includegraphics[width=\columnwidth]{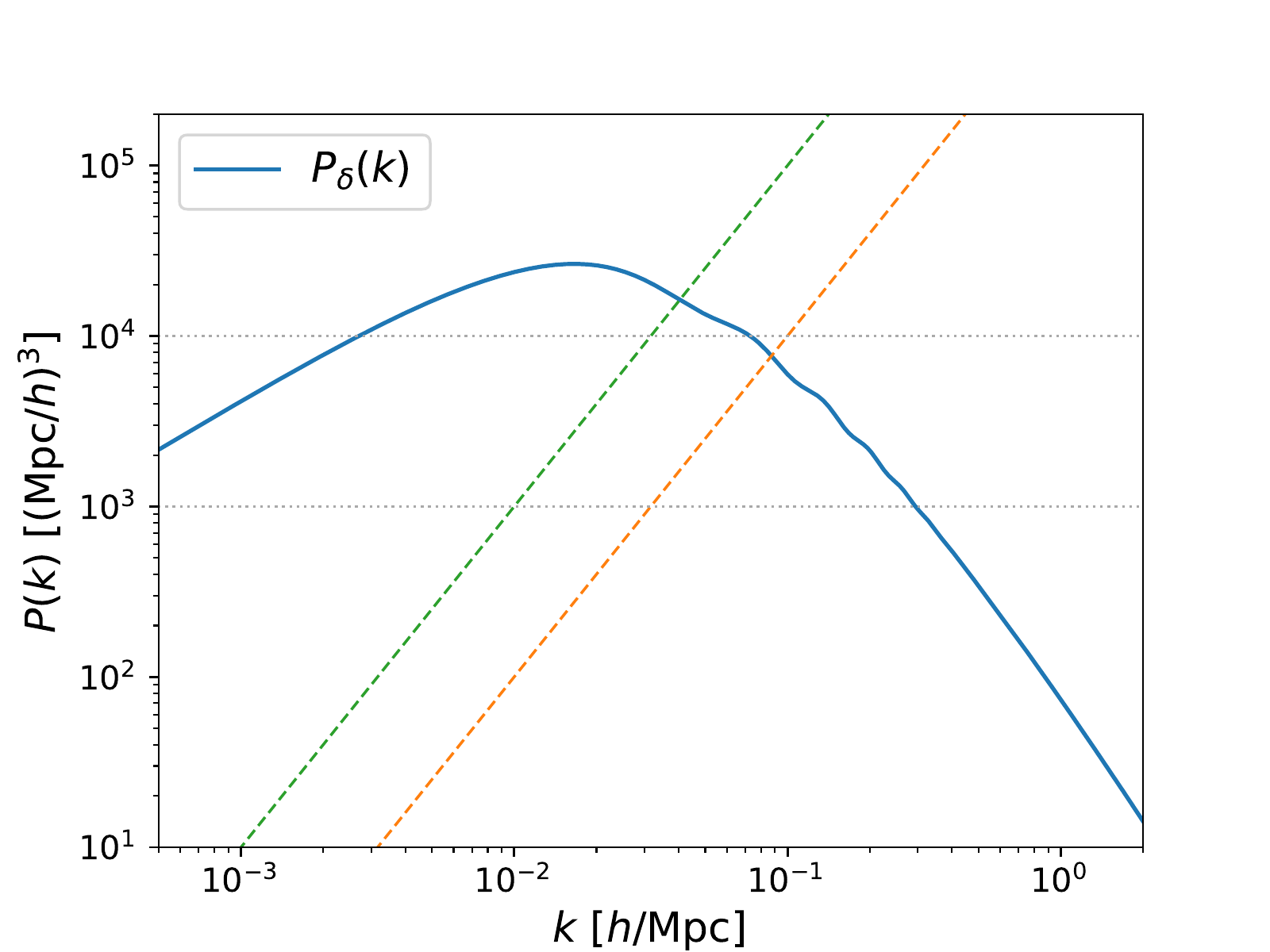}
    \caption{The matter power spectrum in the $\Lambda$CDM cosmology. The dotted line shows the shot noise levels for number densities $\bar{n}=10^{-3}$ and $10^{-4}\,(\mr{Mpc}/h)^{-3}$. The dashed line shows the velocity noise level in the radial direction, i.e. $\bmk=k\hat{\bm{z}}$, $k^2/(afH)^2N_v=k^2\times10^6,k^2\times10^7\,(\mr{Mpc}/h)^5$, which roughly correspond to combining galaxy surveys with number density $\bar{n}=10^{-3}$ and $10^{-4}\,(\mr{Mpc}/h)^{-3}$ with a Simons Observatory-like CMB experiment, respectively.}
    \label{fig:pk3d}
\end{figure}

In this section we consider the reconstruction problem in 3D space.
It is straight forward to generalize the above results to the three dimensions.
The relations are similar to the 1D results, except that the vector field in 3D space has three degrees of freedom.
From the linear continuity equation, we have the theoretical model relating the density and displacement or velocity fields.
We have the density-displacement cross correlation function
\begin{equation}
    \langle\delta(\bmx_j)\psi_\mu(\bmx)\rangle=\int\frac{d^3k}{(2\pi)^3}\frac{-ik_\mu}{k^2}P_\delta(\bmk)\exp{(i\bmk\cdot(\bmx_j-\bmx))},
\end{equation}
and the matter density correlation function is given by
\begin{equation}
\label{eq:corr_dp}
    \langle\delta(\bmx_i)\delta(\bmx_j)\rangle=\int\frac{d^3k}{(2\pi)^3}P_\delta(\bmk)\exp{(i\bmk\cdot(\bmx_i-\bmx_j))}.
\end{equation}
Note that the data covariance includes a shot noise contribution
\begin{equation}
\label{eq:corr_dd}
    \langle\hat{\delta}(\bmx_i)\hat{\delta}(\bmx_j)\rangle=\langle\delta(\bmx_i)\delta(\bmx_j)\rangle+\frac{\delta^D(\bmx_i-\bmx_j)}{\bar{n}(\bmx_i)}.
\end{equation}
Fig.~\ref{fig:pk3d} shows the linear matter power spectrum in the 3D $\Lambda$CDM cosmology at redshift $z=0$, computed using the linear Boltzmann code \verb+CLASS+ \citep{2011JCAP...07..034B}.
We also plot the shot noise levels for $\bar{n}=10^{-3}$ and $10^{-4}\,(\mr{Mpc}/h)^{-3}$, roughly corresponding to the SDSS main sample and eBOSS number densities.
The shot noise dominates over the signal at $k\simeq 0.1$, $0.4\,h\,\mr{Mpc}^{-1}$ for these two number densities, respectively.

To compute the the residual variance in three dimensional Universe, we consider a box of side length $500\,\mr{Mpc}/h$ on a uniform grid with $N^3=256^3$ points. 
Given any position $\bmx$, we can compute the density-displacement cross correlation between $\bmx$ and the data point $\bmx_i$ ($i=1,\dots,N^3$).
For fixed $\bmx$, the cross correlation $\langle{\delta(\bmx_j)\psi_\mu(\bmx)}\rangle$ can be viewed as a vector in $N^3=512^3$ dimensions.
The data covariance $\langle{\hat{\delta}(\bmx_i)\hat{\delta}(\bmx_j)}\rangle$ can be calculated between these $N^3$ data points, which is a $N^3\times N^3$ matrix.
Note that the data points $\bmx_i$ ($i=1,\dots,N^3$) are within the cubic box while $\bmx$ can be any position in the space.
We want to apply the inverse covariance matrix $\langle{\hat{\delta}(\bmx_i)\hat{\delta}(\bmx_j)}\rangle^{-1}$ to the cross correlation vector $\langle{\delta(\bmx_j)\psi_\mu(\bmx)}\rangle$.
The direct operation is very expensive and prohibitive for data with dimensions $N^3\sim10^7$.
Therefore, we instead solve the linear algebra problem $Ax=b$ with the fast approximation method.
We use the preconditioned conjugate gradient method to compute the matrix inversion operation on the cross correlation vector \citep[see e.g.][]{shewchuk1994introduction,2003NewA....8..581P,2012MNRAS.427.2132P}.
We take the preconditioner $M$ as the Fourier transform of $1/P_{\hat{\delta}}$ and the iteration converges in $\mathcal{O}(5)$ steps with the termination criterion $r^TM^{-1}r/ b^TM^{-1}b<10^{-6}$, where $r=b-Ax$ is the residual vector.
In our example, both the density and velocity noises are constant in the observed region and the only irregularities arise from the survey boundary where the noises are infinite outside the survey volume.

\begin{figure}
	\includegraphics[width=\columnwidth]{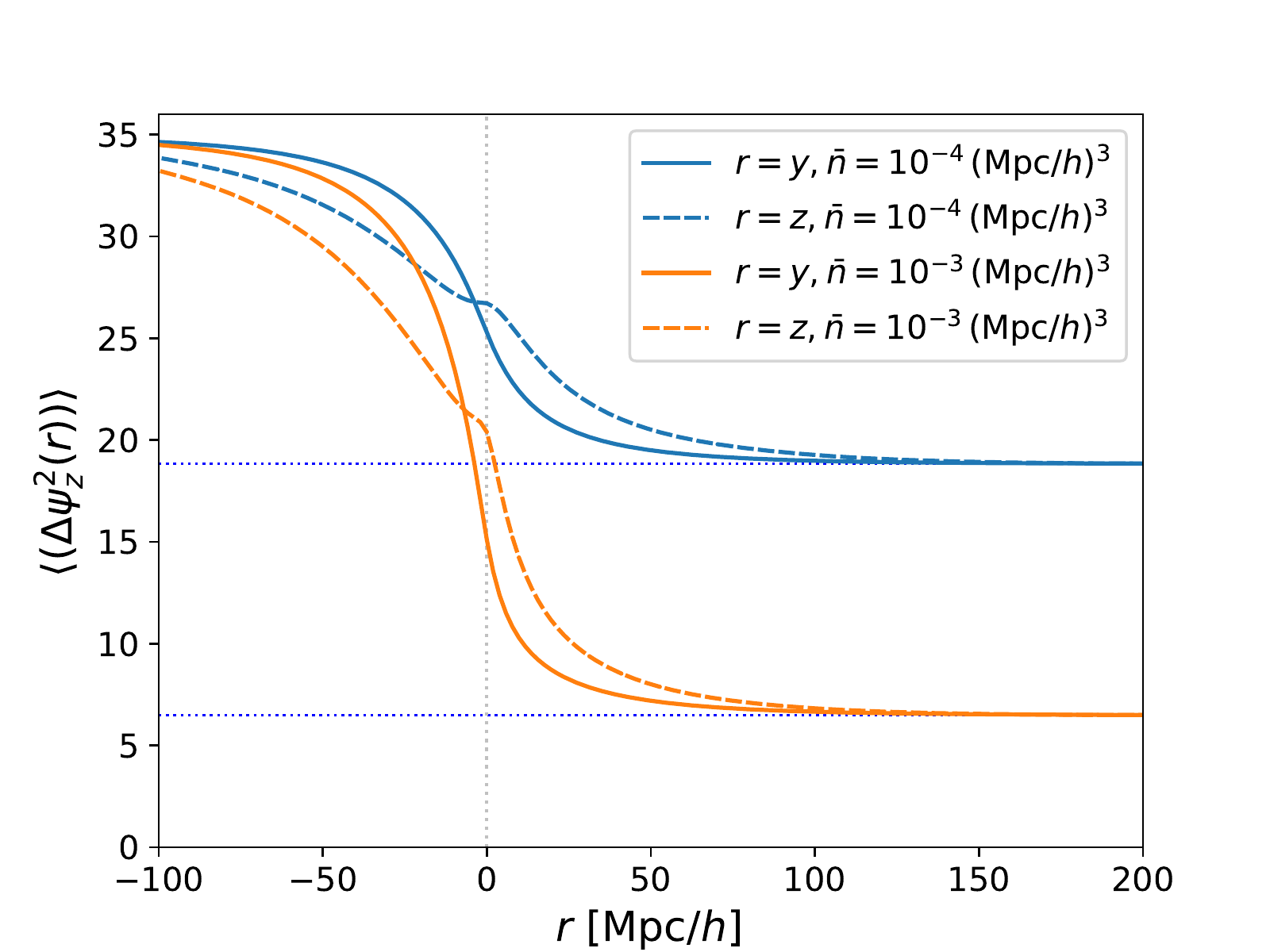}
    \caption{The residual variance $\langle\Delta\psi_z^2(r)\rangle$ for the reconstructed displacement from densities in the 3D space. The residual variance depends on the distance to survey boundary and on whether the line-of-sight direction is perpendicular to or parallel to the boundary. We show results for $\bar{n}=10^{-3}h^3\ \mr{Mpc}^{-3}$ and $\bar{n}=10^{-4}h^3\ \mr{Mpc}^{-3}$.}
    \label{fig:v_pd_F}
\end{figure}

The vector field has three components $(x,y,z)$ in 3D space.  We make the plane-parallel approximation and, without loss of generality, consider the line-of-sight direction to be $z$.  Our focus will be on the radial displacement, $\psi_z$.
Using the linear power spectrum in our fiducial $\Lambda$CDM cosmology the 1D rms displacement is $\langle{\psi_z^2}\rangle=34.89\,(\mr{Mpc}/h)^2$.
The reconstruction error for $\psi_z$ is different when $\psi_z$ is perpendicular to the boundary or parallel to the boundary.
We compute the residual displacement variance for $\bmx=(L/2,y,L/2)$ and $\bmx=(L/2,L/2,z)$.
In the former configuration the displacement vector is parallel to the boundary where $y$ is the distance to the $x-z$ plane and in the latter configuration the displacement is perpendicular to the boundary where $z$ is the distance to the $x-y$ plane. 
Fig.~\ref{fig:v_pd_F} shows the results for different shot noise levels for the two configurations.
The effects of boundary extend to $\sim100\,\mr{Mpc}/h$ in the survey volume.
For the higher number density, the residual variance is increased by $33\%$ and $15\%$ at $y=20\,\mr{Mpc}/h$ and $40\,\mr{Mpc}/h$, for the case where the displacement vector is parallel to the boundary.
When the displacement is perpendicular to the survey boundary the effect is more important, with the residual variance increased by $70\%$ and $33\%$ at $z=20\,\mr{Mpc}/h$ and $40\,\mr{Mpc}/h$, respectively.
The boundary effects are less prominent when the shot noise is high; for $\bar{n}=10^{-4}\,(h/\mr{Mpc})^{-3}$ the increase of the residual variance is $11\%$ and $5\%$ for $y=20\,\mr{Mpc}/h$ and $40\,\mr{Mpc}/h$, and $23\%$ and $12\%$ at $z=20\,\mr{Mpc}/h$ and $40\,\mr{Mpc}/h$, respectively.

The SDSS main sample survey observed galaxies to redshift $z\sim0.2$, i.e. comoving distance $\sim500\,\mr{Mpc}/h$.  The boundary effect degrades the reconstruction performance for regions with $r\lesssim40\,\mr{Mpc}/h$.
From the above estimation, about one third of the total volume is affected by the boundary.
For future low redshift dense surveys like DESI BGS, with higher number density, we expect that the boundary effects will be more important and the velocities from other probes like the kSZ effect can help the reconstruction.
For high redshift surveys like SDSS BOSS, DESI LRG and ELG surveys, the boundary effects are less prominent due to the larger volumes and lower number densities.
However, the reconstructed displacement is usually noisy due to the high shot noise, so velocity information can still improve reconstruction by reducing the reconstruction errors.

In this paper, we consider velocity information from the kSZ tomography \citep{2018arXiv181013423S}, where the larger-scale radial velocity can be obtained by cross-correlating CMB observations and galaxy surveys. 
In linear theory, the radial velocity is related to the matter density as
\begin{equation}
    v_z(\bmk)=\frac{ik_z}{k^2}afH\delta(\bmk).
\end{equation}
The radial velocity inferred from the kSZ effect can be converted into a reconstruction of the matter density field
\begin{equation}
    \hat{\delta}(\bmk)=(ik_z)^{-1}k^2/(afH)\hat{v}_z(\bmk).
\end{equation}
Since the reconstruction noise $N_{v_z}(k)$ approaches a constant on large scales \citep{2018arXiv181013423S}, the reconstruction noise of the density field is
\begin{equation}
    N_\delta(\bmk)=k_z^{-2}k^4/(afH)^2N_{v_z}.
\end{equation}
In Fig.~\ref{fig:pk3d}, we plot the density noise $N_\delta(\bmk)$ for the wave vector $\bmk$ in the $z$ direction, i.e. $\bmk=k\hat{\bm{z}}$, and  $N_\delta(k)=k^2/(afH)^2N_{v_z}=k^2\times10^6,k^2\times10^7\,(\mr{Mpc}/h)^5$. 
Following \citet{2018arXiv181013423S,2018arXiv181013424M}, these velocity reconstruction noise levels correspond roughly to what can be achieved with a CMB experiment with white noise of 6 $\mu$K-arcmin and a $1.5$ arcmin beam, together with a number density of $\bar{n} = 10^{-3}$ and $10^{-4}\,(h/\rm{Mpc})^3$, respectively.
The density modes can be measured better with the kSZ effect than the galaxy surveys on large scales as the noise scales as $k^2$ and approaches zero for $k\to0$.

In three dimensions, the velocity-displacement cross correlation has the form
\begin{equation}
    \langle{v}_\mu(\bmx_j)\psi_\mu(\bmx)\rangle=\int\frac{d^3k}{(2\pi)^3}\frac{afHk_\mu^2}{k^4}P_\delta(\bmk)\exp{(i\bmk\cdot(\bmx_j-\bmx))},
\end{equation}
and the velocity correlation function is
\begin{equation}
    \langle{v}_\mu(\bmx_i)v_\mu(\bmx_j)\rangle=\int\frac{d^3k}{(2\pi)^3}\frac{(afHk_\mu)^2}{k^4}P_\delta(\bmk)\exp{(i\bmk\cdot(\bmx_i-\bmx_j))}.
\end{equation}
The data covariance also includes the measurement noise
\begin{equation}
    \langle{\hat{v}_\mu(\bmx_i)\hat{v}_\mu(\bmx_j)}\rangle=\langle{{v}_\mu(\bmx_i){v}_\mu(\bmx_j)}\rangle+N_{v_\mu}(\bmx_i)\delta^D(\bmx_i-\bmx_j),
\end{equation}
where the velocity noise $N_{v_\mu}(\bmx_i)$ is independent since CMB noise is uncorrelated from one galaxy to the next.

\begin{figure}
	\includegraphics[width=\columnwidth]{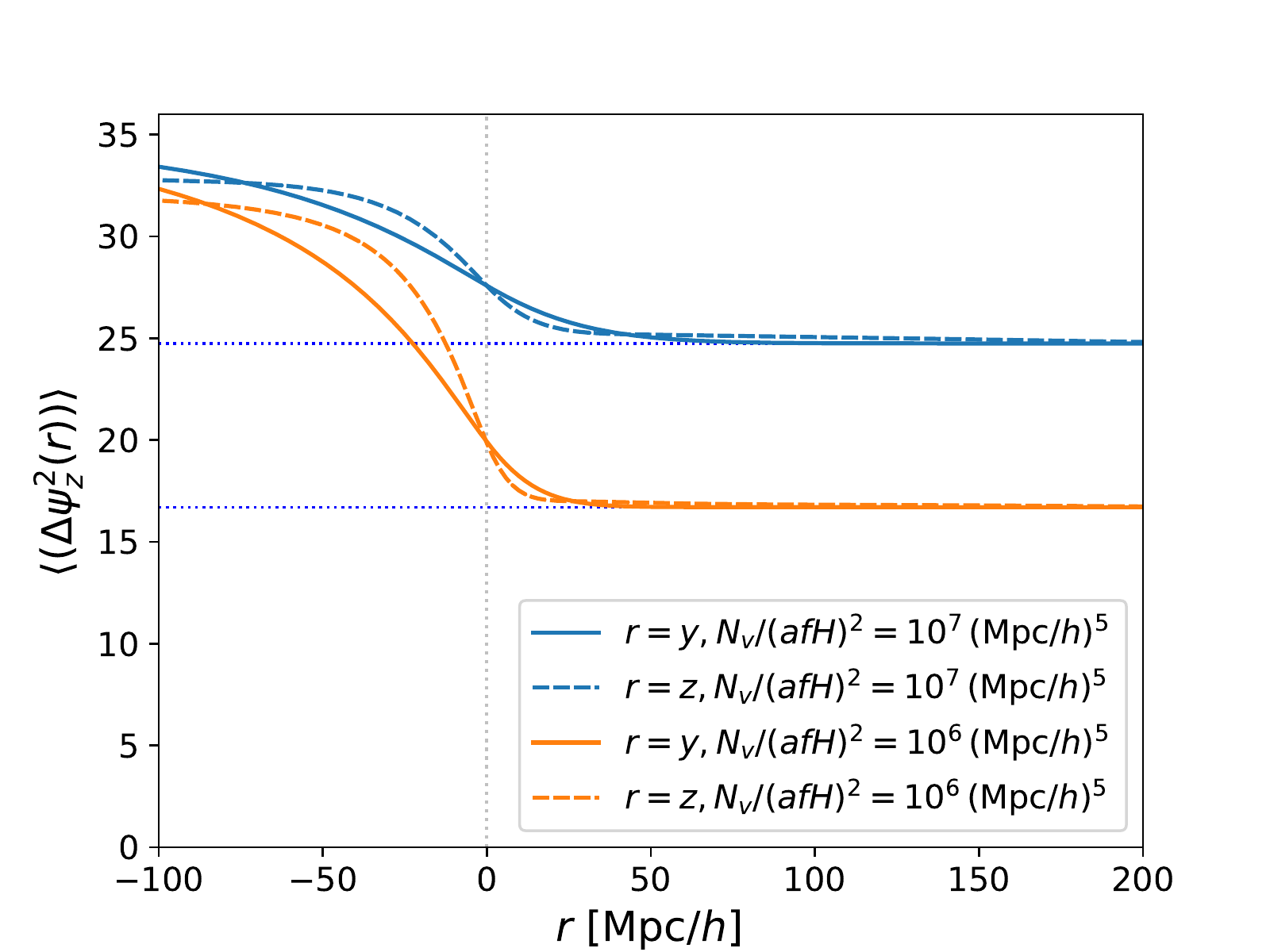}
    \caption{The residual variance $\langle\Delta\psi_z^2(r)\rangle$ for the reconstructed displacement from velocities in the 3D case. The reconstruction is less susceptible to the boundary. The velocity measurement noises are $N_v/(afH)^2=10^6\,(\mr{Mpc}/h)^5$ and $N_v/(afH)^2=10^7\,(\mr{Mpc}/h)^5$.}
    \label{fig:v_pv_F}
\end{figure}

Fig.~\ref{fig:v_pv_F} shows the residual variance of reconstruction with only velocities for different noise levels with two configurations.
The salient feature is that the boundary effects are limited to the region with $r<40\,\mr{Mpc}/h$.
The residual variance is increased by only $20\%$ even at the boundary $r=0\,\mr{Mpc}/h$ for $N_v/(afH)^2=10^6\,(\mr{Mpc}/h)^5$.
For the higher velocity noise, the increase of the residual variance is about $10\%$ for points right on the boundary $r=0\,\mr{Mpc}/h$.
We find that both configurations $r=y$ and $r=z$ show similar behaviours when approaching $r=0\,\mr{Mpc}/h$, while for reconstruction with densities different configurations have very different features near the boundary.
This is because the radial velocity and displacement are related by a simple linear relation determined by the value of $afH$.
Therefore, whether the velocity and displacement vectors are perpendicular or parallel to the boundary does not affect the reconstruction results much.
Notice the errors on reconstruction with velocities are often larger than reconstruction with the density field.
The boundary effect is more prominent than for the 1D results shown in Fig. \ref{fig:v_pp_1d}.
This is because the reconstructed displacement field is the Wiener filtered velocity field, which involves nearby velocity measurements. 
Without measurement noise, we can have perfect displacement reconstruction from velocities near the boundary as we see in the 1D results.
When the noise is large, the reconstruction also depends on the boundary since the Wiener filtering infers the displacement from the measured velocities near this position.

\begin{figure}
	\includegraphics[width=\columnwidth]{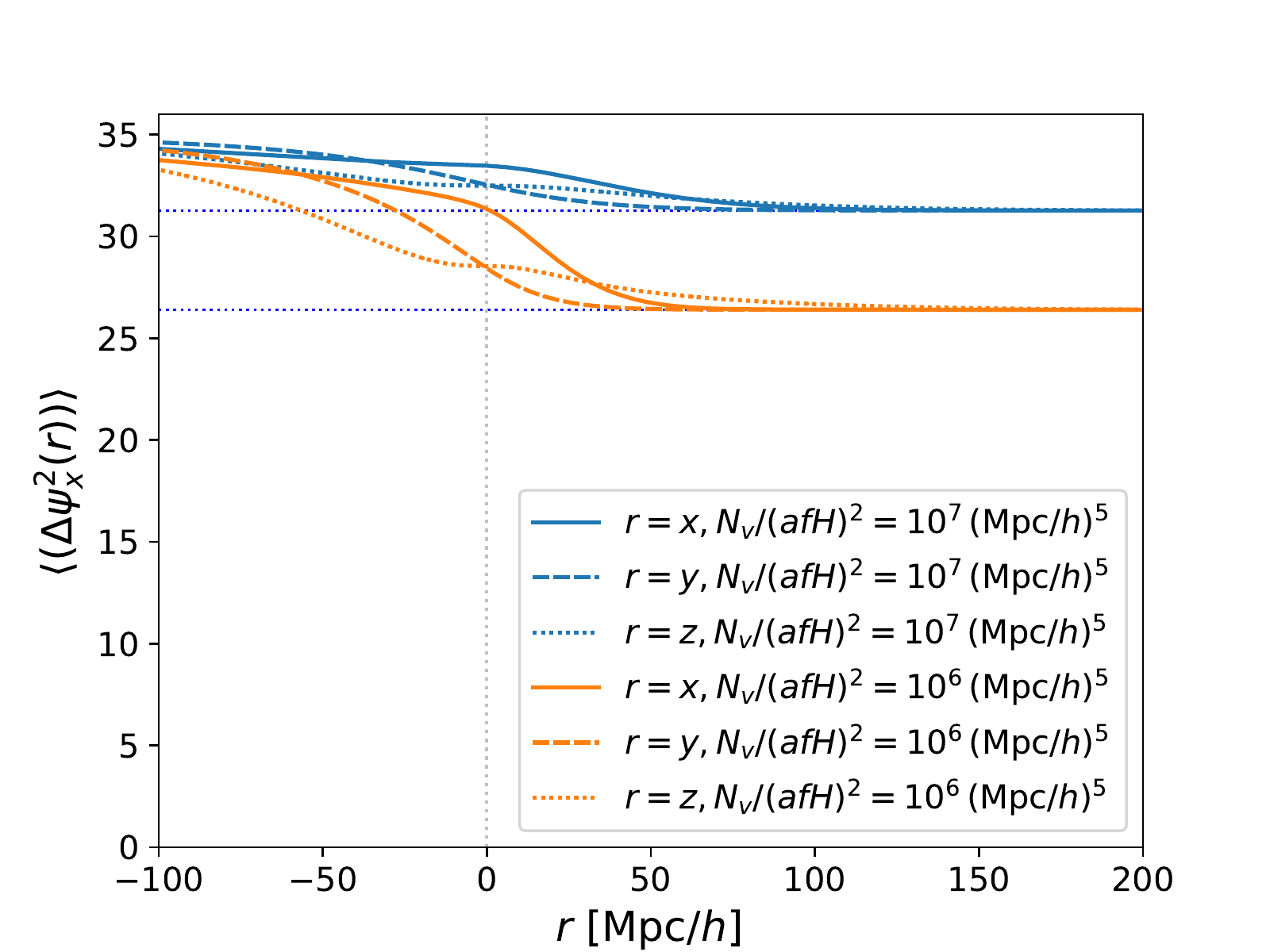}
    \caption{The residual variance $\langle\Delta\psi_x^2(r)\rangle$ for reconstructing $\psi_x$ from $v_z$. The reconstruction is more noisy than reconstruction in the same direction. The velocity measurement noises are $N_v/(afH)^2=10^6\,(\mr{Mpc}/h)^5$ and $N_v/(afH)^2=10^7\,(\mr{Mpc}/h)^5$.}
    \label{fig:v_pv_xz_F}
\end{figure}

The correlation between the radial velocity and velocities in other directions allows us to reconstruct $\psi_x$ and $\psi_y$ from the observation of the radial velocity $v_z$. 
The velocity-displacement cross correlation is then
\begin{equation}
    \langle{v}_\mu(\bmx_j)\psi_\nu(\bmx)\rangle=\int\frac{d^3k}{(2\pi)^3}\frac{afHk_\mu k_\nu}{k^4}P_\delta(\bmk)\exp{(i\bmk\cdot(\bmx_j-\bmx))}.
\end{equation}
Fig.~\ref{fig:v_pv_xz_F} shows the residual variance $\langle{\Delta\psi_x^2(r)}\rangle$ for reconstruction of $\psi_x$ from the radial velocity $v_z$. 
The results are much more noisy than reconstruction in the same direction, i.e. $\psi_z$ from $v_z$.
The residual variance is higher than the latter by $6.53\,(\mr{Mpc}/h)^2$ and $9.70\,(\mr{Mpc}/h)^2$ for the high and low noises, respectively. 
This is due to the correlation between vector fields in different directions is generally weaker than in the same direction.
For the higher velocity noise, the increase of residual variance at $r=0\,\mr{Mpc}/h$ is less than $7\%$.
When the velocity noise is smaller, the variance for $\psi_x$ in the $x$-direction is higher by $18\%$ at $r=0\,\mr{Mpc}/h$, while in the other two directions it is only $8\%$.
To obtain a reliable estimate of the displacement in other directions, we need a better measurement of the radial velocities, although the boundary effect is small.

\begin{figure}
	\includegraphics[width=\columnwidth]{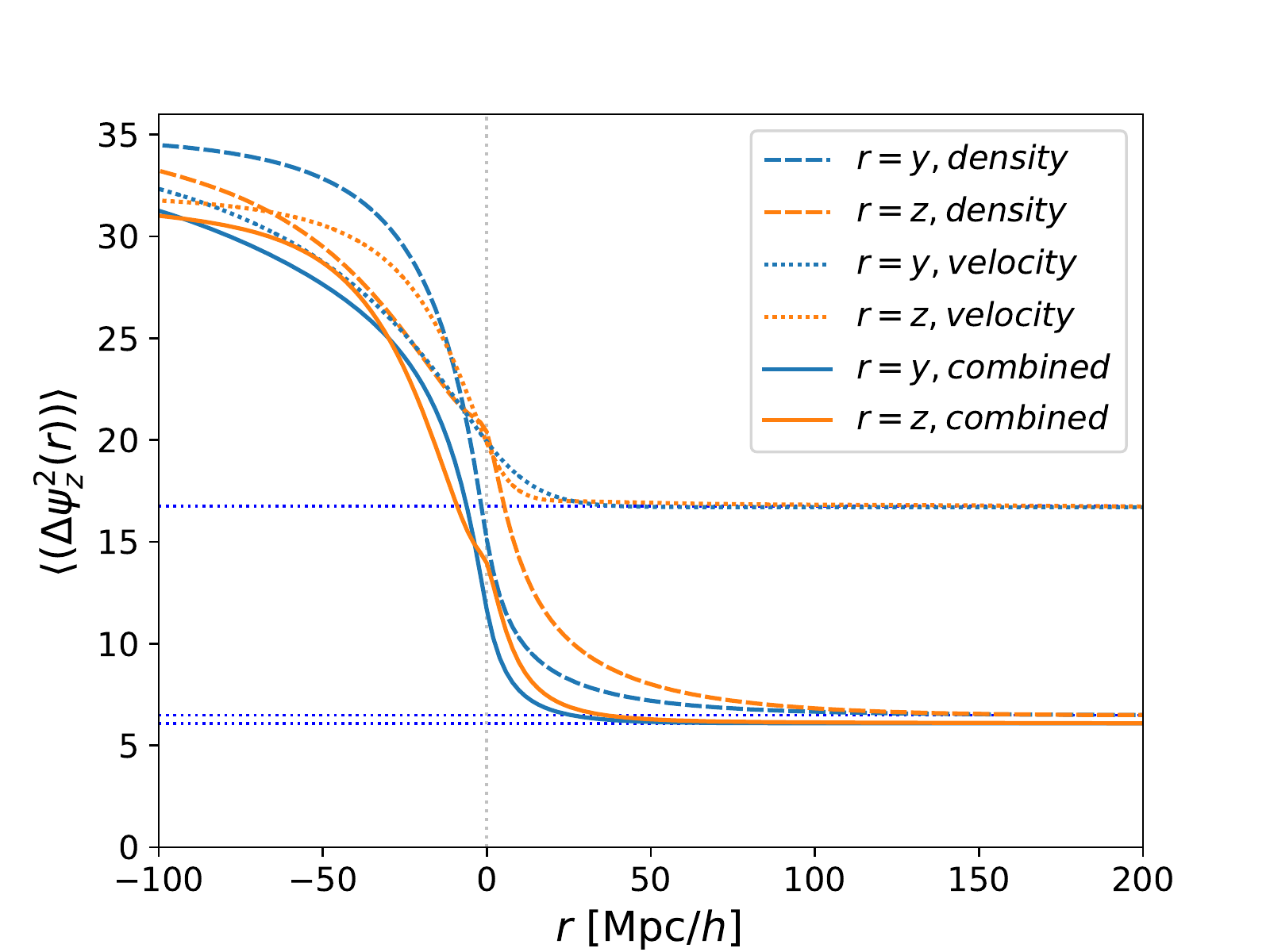}
    \caption{The residual variance $\langle\Delta\psi_z^2(r)\rangle$ for $\bar{n}=10^{-3}\,(\mr{Mpc}/h)^{-3}$ and $N_v/(afH)^2=10^6\,(\mr{Mpc}/h)^5$. The improvement within the survey volume is marginal, but the residual variance near the survey boundary is reduced significantly by including the velocity information and the boundary effects are reduced substantially to $r<25$\,Mpc$/h$ from $r<100$\,Mpc$/h$ with only densities.}
    \label{fig:v_pvd_F}
\end{figure}

Fig.~\ref{fig:v_pvd_F} shows the results by combining both fields with the density noise $\bar{n}=10^{-3}\,(\mr{Mpc}/h)^{-3}$ and velocity noise $N_v/(afH)^2=10^6\,(\mr{Mpc}/h)^5$.
We see that the velocity field is noisy and the residual variance is about two times larger than using the density information only.
Although within the survey volume the improvement is marginal, the residual variance near the survey boundary is reduced significantly by including the velocity information and the boundary effects are reduced to $r<25$\,Mpc$/h$ from $r<100$\,Mpc$/h$.

\begin{figure}
	\includegraphics[width=\columnwidth]{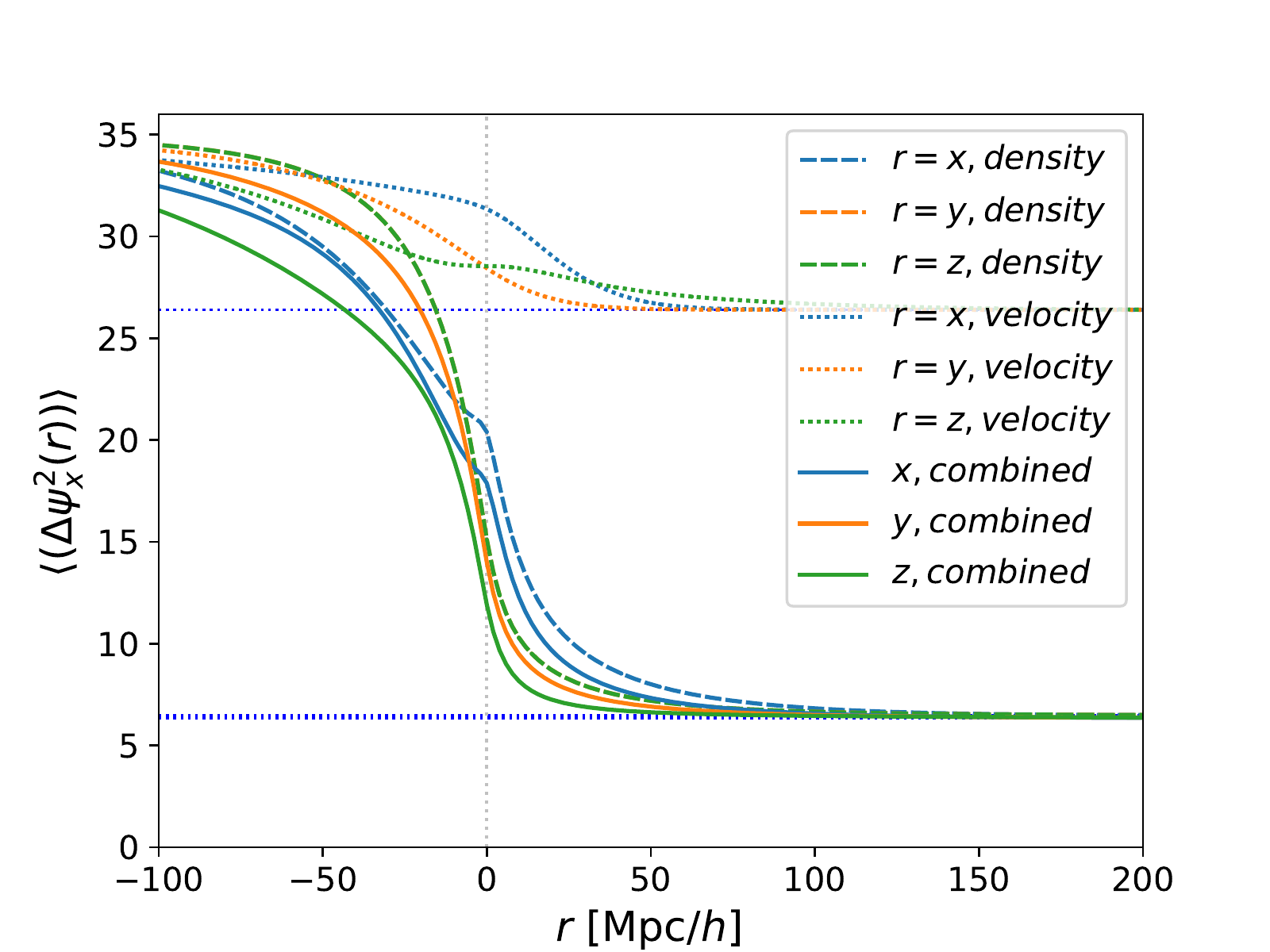}
    \caption{The residual variance $\langle\Delta\psi_x^2(r)\rangle$ for $\bar{n}=10^{-3}\,(\mr{Mpc}/h)^{-3}$ and $N_v/(afH)^2=10^6\,(\mr{Mpc}/h)^5$. The improvement within the survey volume saturates, but the  residual variance near the survey boundary is still reduced by including the velocity information and the boundary effects are reduced to $r\sim50$\,Mpc$/h$.}
    \label{fig:v_pvd_px_F}
\end{figure}

Fig. \ref{fig:v_pvd_px_F} shows the residual variance for $\psi_x$ using the radial information $v_z$.
Note that for the velocity in $x$-direction, the $r=y$ and $r=z$ curves overlap due to the azimuthal symmetry in the $x$-direction.
The improvement within the survey also saturates since the weaker correlation between the vector fields in different directions leads to a noisy reconstruction of $\psi_x$, but the velocity information still helps reconstruction near the boundary, reducing the boundary effects to $r\sim50\,\mr{Mpc}/h$.

\begin{figure}
	\includegraphics[width=\columnwidth]{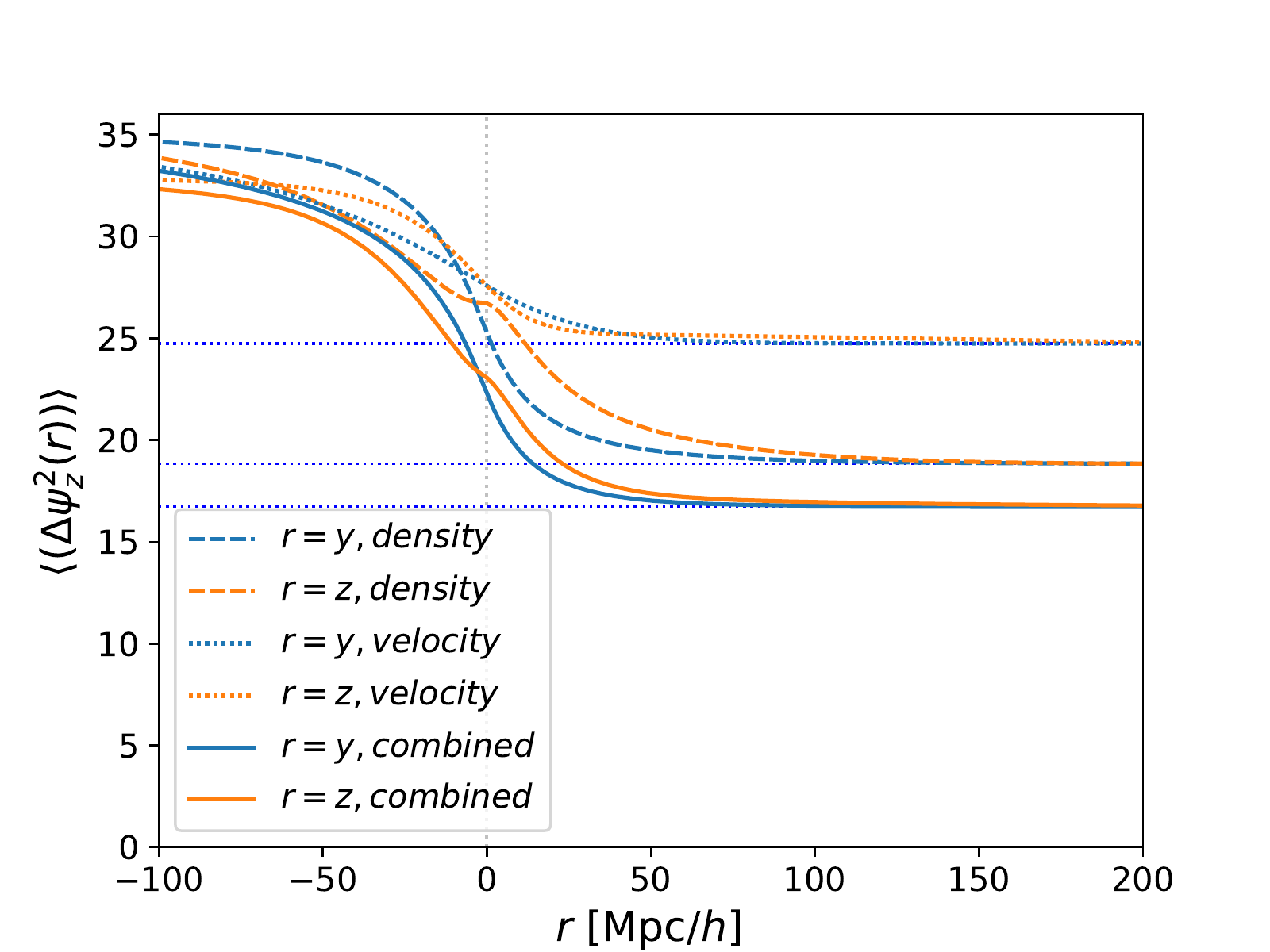}
    \caption{The residual variance $\langle\Delta\psi_z^2(r)\rangle$ for $\bar{n}=10^{-4}\,(\mr{Mpc}/h)^{-3}$ and $N_v/(afH)^2=10^7\,(\mr{Mpc}/h)^5$.
    The residual variance is reduced by $11\%$ within the survey volume by including velocities and the boundary effects are limited to $r\sim15$\,Mpc$/h$ from $r<100$\,Mpc$/h$ with only densities.}
    \label{fig:v_pvd_F_2}
\end{figure}

\begin{figure}
	\includegraphics[width=\columnwidth]{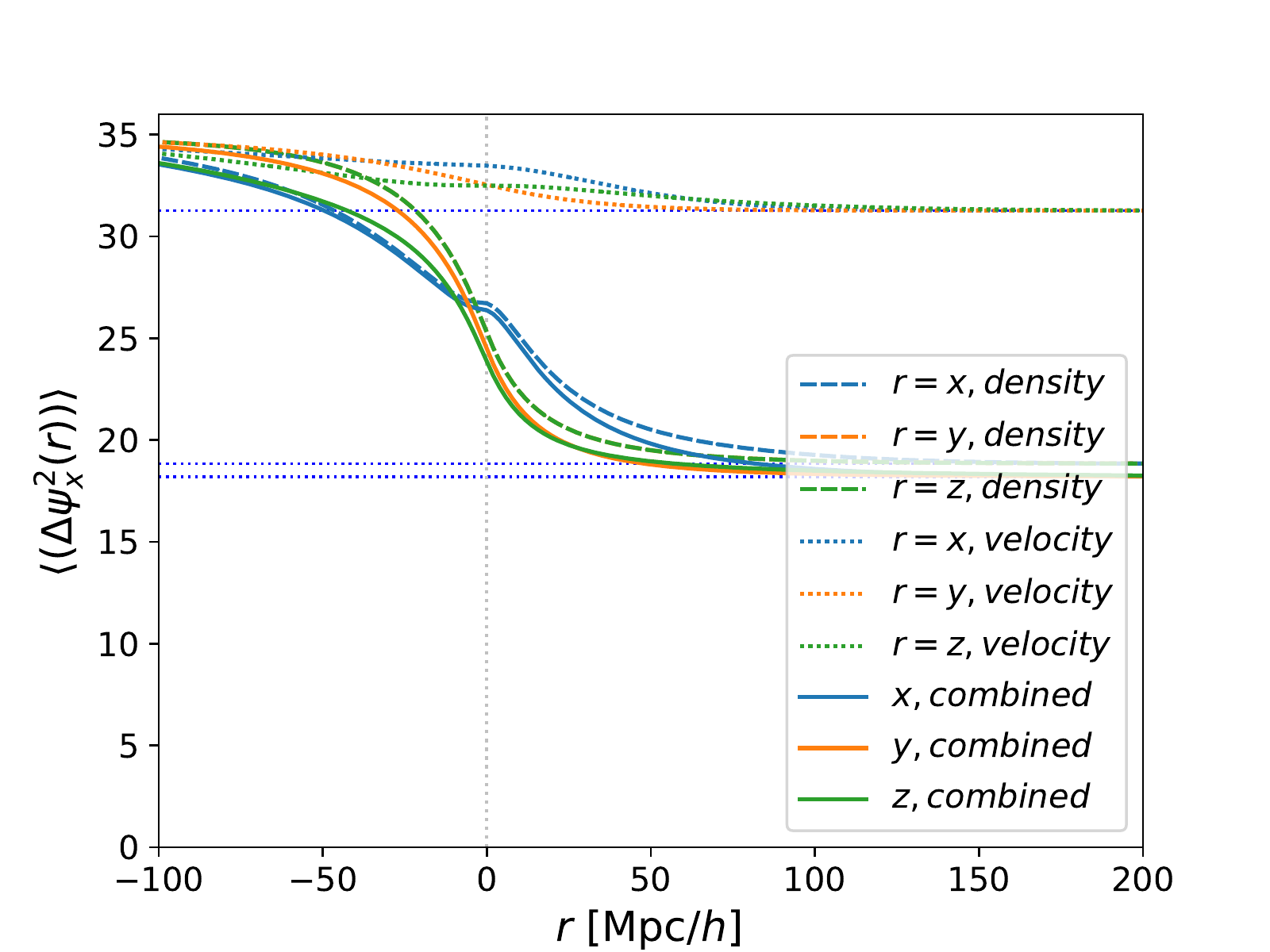}
    \caption{The residual variance $\langle\Delta\psi_x^2(r)\rangle$ for $\bar{n}=10^{-4}\,(\mr{Mpc}/h)^{-3}$ and $N_v/(afH)^2=10^7\,(\mr{Mpc}/h)^5$. The improvement within the survey volume saturates, but the residual variance near the survey boundary is still reduced by including the velocity information and the boundary effects are reduced to $r<50$\,Mpc$/h$.}
    \label{fig:v_pvd_px_F_2}
\end{figure}

Fig. \ref{fig:v_pvd_F_2} and Fig. \ref{fig:v_pvd_px_F_2} show the results for the combination of fields with the density noise $\bar{n}=10^{-4}\,(\mr{Mpc}/h)^3$ and velocity noise $N_v/(afH)^2=10^7\,(\mr{Mpc}/h)^5$.
The residual variance for $\psi_z$ is already reduced by $11\%$ even within the survey volume and the residual variance near the boundary is also reduced substantially near the survey boundary.
Within the range $r>15\,\mr{Mpc}/h$, the residual variance is always smaller than the value within the survey when using densities only.
When reconstructing $\psi_x$ from $v_z$, the improvement saturates within the volume.
However, we still get improvement near the boundary and the effects of boundary is reduces to $r<50\,\mr{Mpc}/h$.

\begin{figure}
	\includegraphics[width=\columnwidth]{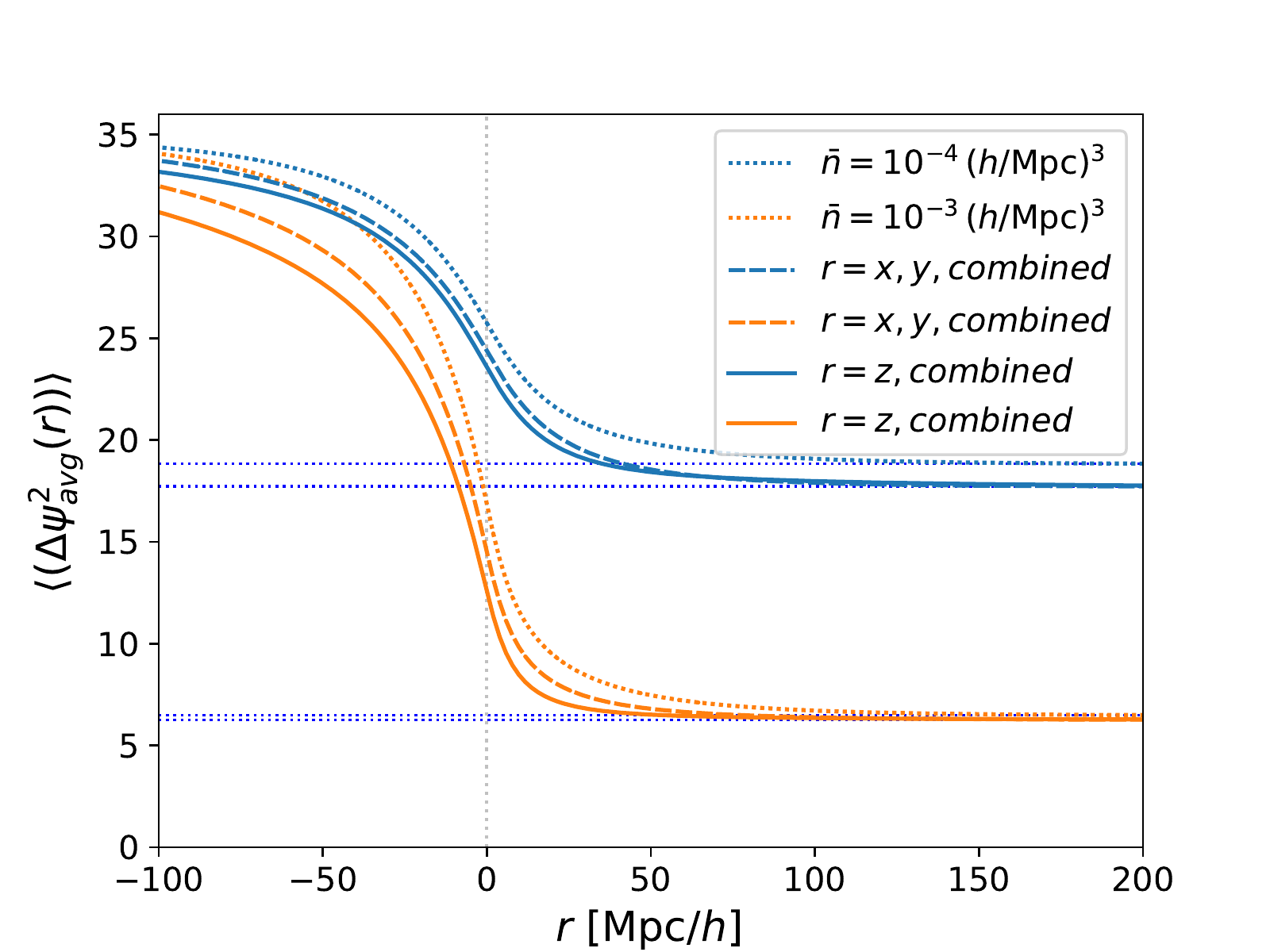}
    \caption{The averaged residual variance $\langle\Delta\psi_{\mr{avg}}^2(r)\rangle$ for three displacement components $\psi_x,\psi_y$ and $\psi_z$, which quantifies the overall performance.
    The boundary effects are reduced from $r\lesssim100\,\mr{Mpc}/h$ to $r\lesssim30\,\mr{Mpc}/h$ and $r\lesssim40\,\mr{Mpc}/h$, respectively, with the velocity information from Simons Observatory.}
    \label{fig:v_pvd_avg}
\end{figure}

Fig. \ref{fig:v_pvd_avg} shows the averaged residual variance for three displacement components in different directions $\psi_x,\psi_y,\psi_z$,
\begin{equation}
    \langle{\Delta\psi_{\mr{avg}}^2(r)}\rangle=\frac{\langle{\Delta\psi_x^2(r)}\rangle+\langle{\Delta\psi_y^2(r)}\rangle+\langle{\Delta\psi_z^2(r)}\rangle}{3},
\end{equation}
where $r=x,y,z$ denotes the different configurations to the boundary.
This quantifies the overall performance by including velocities in the reconstruction.
Note that $N_v/(afH)^2=10^7\,(\mr{Mpc}/h)^5$ and $10^6\,(\mr{Mpc}/h)^5$ are roughly the velocity field noises we can obtain by cross correlating the DESI ELG/LRG survey with effective number density $b^2\bar{n}\sim10^{-4}\,(h/\mr{Mpc})^3$ and SDSS-MGS with $b^2\bar{n}\sim10^{-3}\,(h/\mr{Mpc})^3$ with the CMB observations from the Simons Observatory \citep{2019JCAP...02..056A}. 
We find that the boundary effects are reduced from $r\lesssim100\,\mr{Mpc}/h$ to $r\lesssim30\,\mr{Mpc}/h$ and $r\lesssim40\,\mr{Mpc}/h$, respectively.
This is helpful especially for low redshift high number density surveys like SDSS-MGS and DESI-BGS, where at low redshift the volume is relatively small and a large fraction of the total volume is affected by the boundaries, degrading the performance of reconstruction and thus the BAO measurements.
Although the velocity field measurement are often noisy compared to the density field within the interior of the survey, near the boundary it can substantially improve the reconstruction performance.
Note that with the CMB-S4 observations \citep{2016arXiv161002743A}, the kSZ measurement with lower noise can further improve the reconstruction performance.

\section{Discussion and conclusion}
\label{sec:cls}

We have investigated the problem of reconstructing displacements for positions near the survey boundary and within the interior of the survey volume.
In the former case, the performance degrades because of incomplete data near the boundary and this boundary effect extends to $\sim100$\,Mpc$/h$.
In the computation of residual variances, we have used the Wiener filtering formalism and assumed the linear theory, i.e. the Zeldovich approximation.
For standard reconstruction \citep{2007ApJ...664..675E}, the recovered linear BAO signal is directly related to the quality of the reconstructed Zeldovich displacement \citep[see e.g.][]{2009PhRvD..79f3523P,2009PhRvD..80l3501N,2015MNRAS.450.3822W,2016MNRAS.457.2068C,2016MNRAS.460.2453S,2019arXiv190700043C}.
While for nonlinear reconstruction algorithms \citep[e.g.][]{2017PhRvD..96l3502Z,2017PhRvD..96b3505S,2018PhRvD..97b3505S,2018MNRAS.478.1866H}, solving the large-scale displacement is the first step in reconstruction, which captures the dominant part of linear BAO signal and is most sensitive to the survey boundary.
The following steps reconstruct the small-scale scale displacement to recover the linear signal in the nonlinear regime.
The power of displacement on small scales is much smaller than on large scales with smaller correlation length and thus not affected by the boundary much.
Therefore, the results presented in this paper captures the essence of Lagrangian reconstruction algorithm and the conclusion should be general for most current reconstruction methods.

We have assumed that the density and displacement are fully correlated, which is only valid in the linear theory.
Small-scale stochasticities which are not correlated with the displacement can be thought of as additional noise in the density and velocity fields.
In this sense the effect of imperfect correlation can be thought as increasing the noise power on small scales.

We have used an isotropic linear power spectrum for the theoretical model when compute the correlation matrices, which is a simplified approximation for realistic galaxy density fields.
Since the power spectra of displacement and velocity fields peaks on quite large scales, the linear part contributes most to the total rms displacement and velocity \citep[see e.g.][for more discussions]{2009PhRvD..79f3523P,2015MNRAS.450.3822W}.
The real galaxy distribution also exhibits the anisotropic clustering due to the redshift space distortions.
In particular, the Fingers of God effect which causes the loss of information in the radial direction should be suppressed with anisotropic filtering \citep[see e.g.][]{2016MNRAS.457.2068C,2018MNRAS.478.1866H}.
These more detailed effects need to be quantified using simulations and simulated mocks which we plan to investigate in the future.

We find that if radial velocity information is available, it helps the reconstruction of the radial displacement significantly, while the displacements in the orthogonal directions still benefit from the radial velocity information due to the correlation between different directions.
With the CMB observations from Simons Observatory, the radial velocity measured with the kSZ effect can reduce the effects of boundary to $r\lesssim30\,\mr{Mpc}/h$ and $r\lesssim40\,\mr{Mpc}/h$, for effective number densities of DESI LRG/ELG like survey $b^2\bar{n}\sim10^{-4}\,(h/\mr{Mpc})^3$ and SDSS-MGS with $b^2\bar{n}\sim10^{-3}\,(h/\mr{Mpc})^3$.
The performance can be further improved with CMB-S4.
The velocities can also be included in the forward modeling methods in the optimization process and further augment the the results. 
We therefore expect the joint analysis of large-scale structure and CMB surveys can substantially improve the measurement of the BAO scale and thus constrain cosmological models.



\section*{Acknowledgements}

We acknowledge Marcelo Alvarez, Mathew Madhavacheril, Moritz {M{\"u}nchmeyer} and Marcel Schmittfull for helpful discussions.
M.W. is supported by the U.S. Department of Energy and by NSF grant number 1713791. S.F. is supported by the Physics Division at Lawrence Berkeley National Laboratory.
E.S. is supported by the Chamberlain fellowship at Lawrence Berkeley National Laboratory.
This work made extensive use of the NASA Astrophysics Data System and of the astro-ph preprint archive at arXiv.org.




\bibliographystyle{mnras}
\bibliography{vel} 







\bsp	
\label{lastpage}
\end{document}